\documentclass[a4paper,11pt]{article}

\usepackage{jheppub} 

\usepackage[T1]{fontenc} 
\usepackage{latexsym,amsmath,amsfonts,amssymb,amsthm,mathrsfs, mathtools}

\usepackage[dvipsnames]{xcolor}
\usepackage{setspace}
\newcommand{\floor}[1]{\left\lfloor #1 \right\rfloor}

\title{Holomorphic quasi-modular bootstrap}
\allowdisplaybreaks

\author{Yiwen Pan,}
\author{and Chenxi Zeng}

\affiliation{School of Physics, Sun Yat-sen University,\\No. 135 Xingangxi Road, Guangzhou, Guangdong, China}

\emailAdd{panyw5@mail.sysu.edu.cn}

\emailAdd{zengchx3@mail2.sysu.edu.cn}

\abstract{
Holomorphic modular bootstrap is an approach to classifying rational conformal field theories making use of the modular differential equations. In this paper we explore its flavored refinement. For a class of chiral algebras, we propose constraints on a special null state, which determine the structure of the algebra, and through flavored modular differential equations and quasi-modularity, completely fix the spectra in both the untwisted and twisted sector. Using the differential equations, we reveal hidden structures among null states of the chiral algebras under the modular group action and translation related to spectral flow.

}

\makeatletter
\gdef\@fpheader{}
\makeatother

\begin{document} 
\maketitle
\flushbottom


\section{Introduction}

Classifying conformal field theories (CFT) is an important and challenging endeavor. Although conformal symmetry in general brings strong constraints on the spectrum and dynamics of a theory, the most promising arena for a classification is in two dimensions, or in higher dimensions with enough amount of supersymmetry. In two dimensions, an important class of CFTs receives most attention, the rational CFTs (RCFTs). An RCFT has finitely many primaries, whose fusion algebra can be determined from the modular property of the holomorphic torus partition functions (characters). Well-known examples include the series of Virasoro minimal models \cite{Belavin:1984vu}, and the Wess-Zumino-Witten models \cite{WESS197195,WITTEN1983422} that carries affine Lie algebra symmetries. Furthermore, cosets of these models offer a vast landscape of RCFTs.

Another approach to the classification of RCFTs, referred to as holomorphic modular bootstrap, is based on another organizing principal. The approach focuses on the so called modular linear differential equations (MLDEs) and their solutions \cite{Mathur:1988rx,Mathur:1988na,Eguchi:1987qd,Chandra:2018pjq,Mukhi:2020gnj,Das:2022bxm,Das:2021uvd,Das:2020wsi,Bae:2020xzl,Bae:2021mej,Duan:2022kxr,Duan:2022ltz,Duan:2022ltz,Gowdigere:2023xnm,Mukhi:2022bte,Das:2023qns}. This approach exploits the modular invariance of the torus partition functions of RCFTs, which demands the characters of primaries to form a vector-valued modular form. Using the Wronskian method, this structure can be encoded in the statement that all characters in the theory must be solutions to a (unflavored) modular differential equation, which takes the form
\begin{equation}
  (D_q^{(N)} + \sum_{r = 1}^{N} \phi_{(2N - 2r)} D_q^{(r)})\operatorname{ch} = 0\ .
\end{equation}
Here $D_q^{(N)}$ is an $N$-th order derivative with respect to the modular parameter $q$ coupled with some combinations of Eisenstein series $E_n(\tau)$, $\phi_{r}$ is a meromorphic modular form of weight $r$. This fact is highly useful, since modular forms are well-studied mathematical objects, often organizing into some finite dimensional spaces with known basis. This allows for a systematic enumeration (based on the order $N$ and the Wronskian index) of such MLDEs, and classification of their solutions. With some additional physical and representation-theoretic insights, physicists can identify potential spectra of RCFTs.

The same type of equations also plays a prominent role in the classification of four dimensional superconformal field theories with an $\mathcal{N} = 2$ supersymmetry. It is shown that any 4d  $\mathcal{N} = 2$ unitary SCFT $\mathcal{T}$ corresponds to a 2d non-unitary chiral algebra (or, a vertex operator algebra) $\mathbb{V}[\mathcal{T}]$ \cite{Beem:2013sza,Song:2017oew,Beem:2014rza,Xie:2016evu,Kiyoshige:2020uqz,Bonetti:2018fqz,Lemos:2014lua}, which serves as an important superconformal invariant in the (partial) classification of 4d $\mathcal{N} = 2$ SCFTs. Under the correspondence, the 4d Schur index $\mathcal{I}$ \cite{Gadde:2011uv,Gadde:2011ik} equals the vacuum character $\operatorname{ch}_0$ of the associated chiral algebra, while some non-local BPS operators correspond to non-vacuum modules with the defect index corresponding to the non-vacuum characters \cite{Cordova:2017mhb,Bianchi:2019sxz,Beem:2017ooy,Nishinaka:2018zwq,Dedushenko:2019yiw,Zheng:2022zkm,2023arXiv230409681L,Pan:2024bne}. The Higgs branch of the 4d theory $\mathcal{T}$ is identified with the associated variety of the $\mathbb{V}[\mathcal{T}]$, and consequently the stress tensor of the chiral algebra must be nilpotent in the Zhu's $C_2$-algebra $R_{\mathbb{V}[\mathcal{T}]}$ \cite{Beem:2017ooy}. In favorable situations, this implies the existence of a null state $|\mathcal{N}_T\rangle$ in $\mathbb{V}[\mathcal{T}]$ and a corresponding MLDE that the (unflavored) vacuum character must satisfy. This equation and its solutions encode some crucial data of the 4d theory, including the $a_\text{4d}, c_\text{4d}$ central charges, and the unflavored indices of supersymmetric non-local operators in 4d. 

The MLDEs in the above two contexts have somewhat different conceptual origin, \emph{i.e.}, modularity and the Wronskian method in the first, and null state in the second. Moreover, the modularity in the former context relies on the rationality of the underlying chiral algebra, while in the second, the chiral algebras are in general non-rational: instead, they belong to the larger class of quasi-lisse chiral algebras \cite{Arakawa:2015jya,Arakawa:2016hkg,Beem:2017ooy}. Simplest examples of quasi-lisse chiral algebras, besides the familiar rational ones, are the admissible Kac-Moody algebras, $\widehat{\mathfrak{so}}(8)_{-2}$, $(\widehat{\mathfrak{e}}_6)_{-3}$, the associated chiral algebras of $\mathcal{N} = 4$ $SU(N)$ super Yang-Mills, etc. Compared with those of the rational chiral algebras, the spectra of general quasi-lisse chiral algebras are more involved, which potentially include non-ordinary modules and logarithmic modules. In particular, non-ordinary modules require flavor refinement, which demands generalization of the above MLDEs to flavored MLDEs in the context of modular bootstrap. Even for ordinary modules having unflavoring limit, flavored characters and flavored MLDEs have much better resolution in discerning different modules, since different modules with different flavored characters may share the same unflavored character. There are also interesting observations that some flavored characters of different chiral algebras may coincide when the flavor fugacities take special values \cite{Grover:2022jrc,Abhishek:2023wzp,Buican:2019huq}.

Although the state $|\mathcal{N}_T\rangle$ is guaranteed to exist in an associated chiral algebra $\mathbb{V}[\mathcal{T}]$ \cite{Beem:2017ooy} and in a general quasi-lisse algebra \cite{Arakawa:2016hkg}, and it appears to play an important role in various scenarios, not much is known about it or its properties. The major characteristic is that it satisfies the equation that implements the nilpotency of the stress tensor,
\begin{equation}
  |\mathcal{N}_T\rangle = L_{-2}^N |0\rangle + c_2, \qquad c_2 \in C_2(\mathbb{V}[\mathcal{T}])  \ ,
\end{equation}
and that $|\mathcal{N}_T\rangle$ should be a null state, to be quotient out in the simple/irreducible algebra $\mathbb{V}[\mathcal{T}]$. It would be very useful if more can be said about this state, which helps construct the corresponding flavored or unflavored MLDE. Given a set of (strong) generators of a type of chiral algebras, if there exists more intrinsic (without directly referencing to $T$ or being a null) property of the state $|\mathcal{N}_T\rangle$, then a flavor-refined modular bootstrap approach is possible by starting with a postulate of the general form of $|\mathcal{N}_T\rangle$, followed by analysis of its flavored MLDE to gain insight into the spectrum. We propose that for the Deligne-Cvitanovi\'c series and quasi-lisse $\widehat{\mathfrak{su}}(2)_k$ algebras, the null state $|\mathcal{N}_T\rangle$ can be defined alternatively by 
\begin{equation}
  J^a_{0}|\mathcal{N}_T\rangle = L_2 |\mathcal{N}_T\rangle = h^i_{n \ge 2} |\mathcal{N}_T\rangle = 0 \ .
\end{equation}

In the preliminary study of the flavored MLDEs, these equations exhibit some rich structure \cite{Zheng:2022zkm,Pan:2023jjw}. The coefficients of a flavored MLDE coming from a null state $|\mathcal{N}\rangle$ are Eisenstein series $E_k \big[\substack{\phi \\ \theta}\big]$, which are quasi-Jacobi instead of modular forms (with respect to a modular group $\Gamma \subset SL(2, \mathbb{Z})$). Therefore, the modular transformation of such an equation is inhomogeneous in modular weight. If we anticipate the flavored characters to enjoy certain modularity like their unflavored limit \cite{Arakawa:2016hkg}, the inhomogeneity must imply additional flavored MLDEs, placing additional constraints on $|\mathcal{N}_T\rangle$ and the algebra. We will simply call such property ``quasi-modularity'' as an emphasis on the quasi-Jacobi nature of the equations. These additional flavored MLDEs correspond to additional null states $|\mathcal{N}'\rangle$ in the algebra, and therefore implies a hidden structure among null states under the modular group $\Gamma$. Schematically, it reads $|\mathcal{N}\rangle \xrightarrow{\gamma \in \Gamma} \sum_{\mathcal{N}'}a_{\mathcal{N}'}|\mathcal{N}'\rangle$, where the formal coefficients $a_{\mathcal{N}'}$ will be worked out explicitly. In this paper we reveal these hidden structures of null states in the infinite series of admissible and integrable $\mathfrak{su}(2)$ Kac-Moody algebra, and some other simple examples. In particular, we will argue that in these theories, for any null state $|\mathcal{N}\rangle$ corresponding to a flavored MLDE, the $S$-transformation of $|\mathcal{N}\rangle$ takes the universal form,
\begin{equation}
  S|\mathcal{N}\rangle = \sum_{\ell\ge 0} \sum_{i_1, ... i_\ell = 1}^{r}  \frac{1}{\ell!} \tau^{h[\mathcal{N}] - \ell} \mathfrak{b}_{i_1} \cdots \mathfrak{b}_{i_\ell} h^{i_1}_1 ... h^{i_\ell}_1 |\mathcal{N}\rangle \ ,
\end{equation}
where $h^i_m$ denotes the Cartan generators of an affine Kac-Moody flavor symmetry, and $\mathfrak{b}_i$ are the conjugate fugacities. Apart from a few exceptions, the equations in the modular orbit together determine all the characters of the algebra. Namely, the state $|\mathcal{N}_T\rangle$ with quasi-modularity completely determines the spectrum.

Besides modular transformations, we also investigate the equations corresponding to a null state $|\mathcal{N}\rangle$ under translation $\sigma$ of the fugacities
\begin{equation}
  \mathfrak{b}_i \xrightarrow{\sigma} \mathfrak{b}_i + n_i \tau, \qquad
  \mathfrak{y} \xrightarrow{\sigma} \mathfrak{y} + \sum_{i,j=1}^{r}K^{ij}n_j \mathfrak{b}_i
  + \sum_{i,j = 1}^{r} \frac{1}{2}K^{ij} n_i n_j \tau \ ,
\end{equation}
We find that, in terms of the corresponding null state,
\begin{align}
  |\mathcal{N}\rangle \xrightarrow{\sigma} \sum_{\ell \ge 0}\sum_{i_1\le ...\le i_\ell} \frac{(-1)^\ell}{\ell!} n_{i_1} ... n_{i_\ell} h^{i_1}_{1} \cdots h^{i_\ell}_{1} |\mathcal{N}\rangle \ .
\end{align}
Such fugacity translation generates new characters from an existing one, and is closely related to the spectral flow operation \cite{Ridout:2008nh,Creutzig:2012sd,Cordova:2017mhb,2023arXiv230409681L,Kawasetsu:2021qls}. Using these translations, we can easily construct twisted flavored MLDEs satisfied by twisted module characters, and find the twisted spectrum by solving all the twisted equations in the modular orbit.

The paper is organized as follows. In section 2 we recall some facts and notations on Kac-Moody algebras and modular differential equations. In section 3 we propose some intrinsic constraints on the state $|\mathcal{N}_T\rangle$ and a procedure to determine the spectrum of an algebra. In section 4 we apply the proposal to a weight-four state $|\mathcal{N}_T\rangle$ and recover the Deligne-Cvitanovi\'c series. In section 5 we apply the proposal to $\widehat{\mathfrak{su}}(2)_k$ algebras to recover all the integrable and admissible spectra. In section 6, we study the action of translation of the flavored fugacities on the flavored characters, the flavored MLDEs and the null states, and construct the twisted flavored MLDEs, which further determine the twisted spectrum.

\section{Flavored modular differential equations\label{section:background}}

\subsection{Quasi-Jacobi forms}

In writing the flavored MLDEs, we will use extensively the (twisted) Eisenstein series $E_k\big[\substack{\phi \\ \theta} \big]$. It is defined as a $q$-series \cite{Mason:2008zzb}
\begin{align}
  E_{k \ge 1}\left[\begin{matrix}
    \phi \\ \theta
  \end{matrix}\right] \coloneqq - \frac{B_k(\lambda)}{k!}  + \frac{1}{(k-1)!}\sum_{r \ge 0}' \frac{(r + \lambda)^{k - 1}\theta^{-1} q^{r + \lambda}}{1 - \theta^{-1}q^{r + \lambda}}
  + \frac{(-1)^k}{(k-1)!}\sum_{r \ge 1} \frac{(r - \lambda)^{k - 1}\theta q^{r - \lambda}}{1 - \theta q^{r - \lambda}} \ , \nonumber
\end{align}
where $\phi = e^{2\pi i \lambda}$ and $\theta$ are often referred to as the characteristics. In this paper we will call $\phi$ the twist parameter as later discussions of twisted modules relates to this parameter. We will also call $k$ the (modular) weight of the Eisenstein series, as it is tied to the transformation property of $E_k$ under $SL(2, \mathbb{Z})$. $B_k(x)$ denotes the $k$-th Bernoulli polynomial, and the prime $^\prime$ means that the term with $r = 0$ should be omitted when $\phi = \theta = 1$. We also define $E_0\big[\substack{\phi\\\theta}\big] = -1$. In the limit $\phi, \theta \to 1$, we recover the standard Eisenstein series $E_k$,
\begin{equation}
	E_{2n}\begin{bmatrix}
		+1 \\ +1
	\end{bmatrix} = E_{2n}(\tau), \quad
	E_{2n + 1 \ge 3} \begin{bmatrix}
		+1 \\ +1
	\end{bmatrix} = E_{2n + 1}(\tau) = 0, \quad
	E_1 \begin{bmatrix}
		+1 \\ z
	\end{bmatrix}
	= \frac{1}{2\pi i } \frac{\vartheta_1'(\mathfrak{z})}{\vartheta_1(\mathfrak{z})} \ . 
\end{equation}
Note that $E_1 \big[\substack{1 \\ z}\big]$ has a simple pole at $z = 1$ (or, $\mathfrak{z} = 0$), since $\vartheta_1(\mathfrak{z}\to 0) \sim 2 \pi \mathfrak{z}q^{1/8}$, but $\vartheta'_1(0)\ne 0$.

The Eisenstein series can be acted on by $SL(2, \mathbb{Z})$, generated by
\begin{align}
  S: \tau \to - \frac{1}{\tau}, \ \mathfrak{z} \to \frac{\mathfrak{z}}{\tau}, \qquad
  \qquad
  T: \tau \to \tau + 1 , \ \mathfrak{z} \to \mathfrak{z} \ .
\end{align}
The Eisenstein series transforms non-trivially. 
Explicitly, $E_k \big[\substack{\pm 1 \\ \pm z}\big]$ transform under $S$,
\begin{align}\label{Eisenstein-S-transformation}
  E_n \begin{bmatrix}
    +1 \\ +z
  \end{bmatrix} \xrightarrow{S} &
  \left(\frac{1}{2\pi i}\right)^n\left[\bigg(\sum_{k \ge 0}\frac{1}{k!}(- \log z)^k y^k\bigg)
  \bigg(\sum_{\ell \ge 0}(\log q)^\ell y^\ell E_\ell \begin{bmatrix}
    + 1 \\ z
  \end{bmatrix}\bigg)\right]_n\ ,\\
  E_n \begin{bmatrix}
    -1 \\ +z
  \end{bmatrix} \xrightarrow{S} &
  \left(\frac{1}{2\pi i}\right)^n\left[\bigg(\sum_{k \ge 0}\frac{1}{k!}(- \log z)^k y^k\bigg)
  \bigg(\sum_{\ell \ge 0}(\log q)^\ell y^\ell E_\ell \begin{bmatrix}
    + 1 \\ -z
  \end{bmatrix}\bigg)\right]_n\ ,\\
  E_n \begin{bmatrix}
    1 \\ -z
  \end{bmatrix} \xrightarrow{S} &
  \left(\frac{1}{2\pi i}\right)^n\left[\bigg(\sum_{k \ge 0}\frac{1}{k!}(- \log z)^k y^k\bigg)
  \bigg(\sum_{\ell \ge 0}(\log q)^\ell y^\ell E_\ell \begin{bmatrix}
    -1 \\ +z
  \end{bmatrix}\bigg)\right]_n\ ,\\
  E_n \begin{bmatrix}
    -1 \\ -z
  \end{bmatrix} \xrightarrow{S} &
  \left(\frac{1}{2\pi i}\right)^n\left[\bigg(\sum_{k \ge 0}\frac{1}{k!}(- \log z)^k y^k\bigg)
  \bigg(\sum_{\ell \ge 0}(\log q)^\ell y^\ell E_\ell \begin{bmatrix}
    -1 \\ -z
  \end{bmatrix}\bigg)\right]_n\ ,
\end{align}
where $[ \ldots ]_n$ extracts the coefficient of $y^n$. Under the $T$-action,
\begin{align}
  E_n \begin{bmatrix}
    + 1 \\ + z
  \end{bmatrix} \xrightarrow{T}& \ E_n \begin{bmatrix}
    + 1 \\ + z
  \end{bmatrix}, & 
  E_n \begin{bmatrix}
    - 1 \\ + z
  \end{bmatrix} \xrightarrow{T}& \
  E_n \begin{bmatrix}
    - 1 \\ - z
  \end{bmatrix} \\
  E_n \begin{bmatrix}
    + 1 \\ - z
  \end{bmatrix} \xrightarrow{T}& \ E_n \begin{bmatrix}
    + 1 \\ - z
  \end{bmatrix}, & 
  E_n \begin{bmatrix}
    - 1 \\ - z
  \end{bmatrix} \xrightarrow{T}& \ 
  E_n \begin{bmatrix}
    - 1 \\ + z
  \end{bmatrix} \ .
\end{align}
We combine the $S, T$ transformations and see that under $STS$,
\begin{align}
  E_n \begin{bmatrix}
    -1 \\ z
  \end{bmatrix} \xrightarrow{STS}
  \left(\frac{1}{2\pi i}\right)^n\left[\bigg(\sum_{k \ge 0}\frac{1}{k!}(- \log z)^k y^k\bigg)
  \bigg(\sum_{\ell \ge 0}(\log q - 2\pi i)^\ell y^\ell E_\ell \begin{bmatrix}
    -1 \\ +z
  \end{bmatrix}\bigg)\right]_n\ . \nonumber
\end{align}

Eisenstein series are special cases of quasi-Jacobi forms. A modular form $f_k(\tau)$ of modular-weight $k$ is a meromorphic function of $\tau$ in the upper half plane, which transforms under $SL(2, \mathbb{Z})$ as
\begin{equation}
	f_k(\frac{a \tau + b}{c \tau + d}) = (c \tau + d)^k f_k(\tau) \ , \qquad \begin{pmatrix}
		a & b \\
		c & d
	\end{pmatrix} \in SL(2, \mathbb{Z}) \ .
\end{equation}
Modular forms can be written as polynomials of Eisenstein series $E_4(\tau), E_6(\tau)$. One can also consider modular forms with respect to congruence subgroups of $SL(2, \mathbb{Z})$, for example, $\Gamma = \Gamma(2), \Gamma_\theta, \Gamma^0(2)$ and $\Gamma_0(2)$ when discussing chiral algebras with fermionic generators \cite{Bae:2020xzl}. These forms are generated by suitable (symmetric combinations of) Jacobi theta functions $\vartheta_i(0)^4$. For example, the $\Gamma^0(2)$ modular forms are spanned by $\vartheta_2(0)^{4r} \vartheta_3(0)^{4s} + \vartheta_3(0)^{4r} \vartheta_2(0)^{4s}$, or in terms of polynomials in Eisenstein series $E_4(\tau), E_6(\tau)$ and $E_k \big[\substack{-1 \\ 1}\big]$.

A generalization of $SL(2, \mathbb{Z})$ modular form is the notion of quasi-modular forms, which are generated by $E_2(\tau), E_4(\tau), E_6(\tau)$. They transform under $SL(2, \mathbb{Z})$ inhomogeneously,
\begin{align}
	f_k(\frac{a \tau + b}{c \tau + d}) = \sum_{\ell = 0}^k (c\tau + d)^{k - \ell} g_\ell(\tau) \ .
\end{align}

The modular transformations of Eisenstein series suggests that they are one-parameter generalization of the quasi-modular form, referred to as quasi-Jacobi forms \cite{2014arXiv1406.1139O,Krauel:2013lra}. In particular, the Eisenstein series $E_k \big[\substack{+1 \\ z}\big]$ that will appear in Zhu's recursion can be written as polynomials in the generators
\begin{align}
	E_1 \begin{bmatrix}
		+1 \\ z
	\end{bmatrix}, \qquad E_2(\tau), \qquad E_4(\tau), \qquad 
	\wp(\mathfrak{z}), \qquad \wp'(\mathfrak{z}) \ .
\end{align}
where $\wp(\mathfrak{z})$ denotes the standard Weierstrass $\wp$ function, and can be computed as
\begin{equation}
	\wp(\mathfrak{z}) = \left({\frac{\vartheta'_1(\mathfrak{z})}{\vartheta_1(\mathfrak{z})}}\right)^2
	- \frac{\vartheta''_1(\mathfrak{z})}{\vartheta_1(\mathfrak{z})} + \frac{1}{3}\frac{\vartheta_1'''(0)}{\vartheta_1'(0)} \ .
\end{equation}
In all of the above types of functions, the modular weight is additive, and it can be increased by $2N$ using the differential operator $D_q^{(N)}$,
\begin{equation}
	D_q^{(N)} \coloneqq \partial_{(2N - 2)} \circ \partial_{(2N - 4)} \circ \cdots \circ \partial_{(2)} \circ \partial_{(0)} \ , \qquad \partial_{(k)} \coloneqq q \partial_q + k E_2(\tau) \ .
\end{equation}
Similarly, under $S: \tau \to - \tau^{-1}, \mathfrak{z} \to \tau^{-1} \mathfrak{z}$, $D_{z} \coloneqq z\partial_{z}$ also increases the modular weight by $1$. These differential operators are crucial ingredients of flavored MLDEs.

\subsection{Kac-Moody algebra}

In this paper we will focus on Kac-Moody algebras $\widehat{\mathfrak{g}}_k$. In this section we will briefly recall some standard notation and well-known results concerning simple Lie algebras and the corresponding Kac-Moody algebras.

Consider a simple Lie algebra $\mathfrak{g}$ of rank $r$. The generators of $\mathfrak{g}$ will be generically denoted as $J^a$ with the commutation relations $[J^a, J^b] = f^{ab}{_c} J^c$ with the structure constants $f^{ab}{_c}$. The roots and simple roots of $\mathfrak{g}$ are denoted as $\alpha$ and $\alpha_i$, and the collection of roots $\Delta \coloneqq \{\alpha\}$. Positive roots will be indicated by $\alpha > 0$. We define the Killing form $K(\cdot , \cdot)$ to be a symmetric bilinear function given by
\begin{align}
	K(X, Y) \coloneqq \frac{1}{2h^\vee} \operatorname{tr}_\text{adj} XY,  \qquad K^{ab} \coloneqq K(J^a, J^b)\ , \qquad \forall J^a, J^b, X, Y \in \mathfrak{g} \ .
\end{align}
We also use the inverse matrix $K_{ab}$ such that $\sum_{b}K_{ab}K^{bc} = \delta_a^c$. Using the Killing form one can define an inner product $(\cdot, \cdot)$ between the weights, such that for any long root $\alpha$, $|\alpha|^2 \coloneqq (\alpha, \alpha) = 2$. The simple coroots are $\alpha_i^\vee \coloneqq 2\alpha_i/|\alpha_i|^2$, dual to the fundamental weights $\omega_i$. Similarly, the dual basis of the simple roots $\alpha_i$ are the fundamental coweights $\omega_i^\vee$, such that $(\alpha_i, \omega^\vee_j) = (\alpha_i^\vee, \omega_j) = \delta_{ij}$. The highest root and the Weyl vector are denoted as $\theta$ and $\rho$, and the dual Coxeter number is defined to be $h^\vee \coloneqq (\theta, \rho) + 1$. The Cartan matrix of $\mathfrak{g}$ is given by $A_{ij} \coloneqq (\alpha_i, \alpha_j^\vee)$.

In concrete computations, we often employ the Chevalley basis of generators. The basis vectors associated to a simple root $\alpha_i$ are denoted by $e^i, f^i, h^i$, with the non-vanishing commutation relations
\begin{align}
	[e^i, f^j] = \delta_{ij} h^j, \qquad
	[h^i, e^j] = A_{ji}e^j, \qquad
	[h^i, f^j] = - A_{ji}f^j \ , \qquad i = 1, \ldots, r \ .
\end{align}
In particular, $\{h^i\}_{i = 1}^r$ span the Cartan subalgebra $\mathfrak{h} \subset \mathfrak{g}$. The remaining ladder operators $E^\alpha$ corresponding to an arbitrary roots $\alpha$ can be constructed by suitable nested commutators of $e$'s and $f$'s, and in particular, $e^i = E^{\alpha_i}$, $f^i = E^{- \alpha_i}$. In the end, all of the $h^i$ and $E^\alpha$ form a complete basis of $\mathfrak{g}$, with the standard commutation relations
\begin{align}
	[E^\alpha, E^{-\alpha}] = & \ \sum_{i = 1}^{r} f^{\alpha, -\alpha}{_i}h^i, \qquad
	[E^\alpha, E^\beta] = f^{\alpha \beta}{_\gamma} E^\gamma \text{ if } \gamma = \alpha + \beta \in \Delta \ , \\
	[h^i, E^\alpha] = & \ (\alpha, \alpha_i^\vee) E^\alpha \ .
\end{align}
When acting on a finite dimensional representation, the eigenvalues of $h^i$ are simply the Dynkin labels $\lambda_i$. Under the Chevalley basis, the Killing form drastically simplifies, and the only non-zero components are
\begin{align}
	K^{ij} \coloneqq & \ K(h^i, h^j) = (\alpha_i^\vee, \alpha_j^\vee), \qquad \text{with} \qquad \sum_{j = 1}^{r} K_{ij}K^{jk} = \delta_i^k, \\
	K^{\alpha, -\alpha} \coloneqq & \ K(E^{\alpha}, E^{-\alpha}), \qquad K_{\alpha, -\alpha} = (K^{\alpha, -\alpha})^{-1} \ .
\end{align}
In particular, for any two weights $\mu, \nu$ the inner product can be computed using $K^{ij}$,
\begin{align}
	(\mu, \nu) = \sum_{i,j = 1}^{r} K^{ij}\mu_i \nu_i, \quad \text{where} \quad \mu = \sum_{i = 1}^{r} \mu_i\alpha_i^\vee, \quad \nu = \sum_{i = 1}^{r} \nu_i \alpha_i^\vee \ .
\end{align}
The structure constants are also related to $K$, 
\begin{align}
	f^{\alpha, -\alpha}{_i} = \frac{|\alpha_i|^2}{2} m^\alpha_i K^{\alpha, -\alpha} \ , \qquad
	f^{i \alpha}{_\alpha} = (\alpha, \alpha_i^\vee) \ , \qquad \text{with} \quad \alpha = \sum_{i = 1}^{r} m^\alpha_i \alpha_i \ .
\end{align}

Now we turn to the Kac-Moody algebra $\widehat{\mathfrak{g}}_k$ corresponding to a simple Lie algebra $\mathfrak{g}$. The generators are denoted as $J^a_n$, and we often take the Chevalley basis $J^a = h^i, E^\alpha$, with the standard commutation relations
\begin{align}
	[J^a_m, J^b_n] = \sum_{c}  f^{ab}{_c} J^c_{m + n} + m k K^{ab} \delta_{m + n , 0} \ .
\end{align}
Here $k$ denotes the central element in the algebra, which can be taken simply as a number called the level. We assemble the generators into local operators $J^a(z) \coloneqq \sum_{n \in \mathbb{Z}} J^a_n z^{-n - 1}$, and the above commutation relations are equivalent to the standard OPE
\begin{align}
	J^a(z) J^b(w) = \frac{\sum_c f^{ab}{_c} J^c(w)}{z - w} + \frac{k K^{ab}}{(z - w)^2} + O(z - w) \ .
\end{align}
When the level is non-critical, namely, $k \ne - h^\vee$, we define the Sugawara stress tensor $T_\text{Sug}(z)$ by the normal-ordered product,
\begin{align}
	T_\text{Sug}(z) = \frac{1}{2(k + h^\vee)} \sum_{A,B} K_{ab} (J^aJ^b)(z) = \sum_{n \in \mathbb{Z}} L_n z^{-n - 2} \ .
\end{align}
The mode $L_n$ can be computed in terms of $J^a_m$,
\begin{align}
	L_{n} = \frac{1}{2(k + h^\vee)} \sum_{a,b} \biggl(\sum_{m \le -1} K_{ab} J^a_m J^b_{n - m} + \sum_{m \ge 0}K_{ab}J^a_{n - m}J^b_m\biggr)  \ .
\end{align}
They satisfy the Virasoro commutation relations with the Sugawara central charge $c$,
\begin{align}
	[L_m, L_n] = (m - n)L_{m + n} + \frac{c}{12}m(m+1)(m - 1)\delta_{m + n, 0}, \qquad
	c = \frac{k \dim \mathfrak{g}}{k + h^\vee} \ .
\end{align}
In other words, away from criticality the affine Kac-Moody algebra $\widehat{\mathfrak{g}}_k$ contains a Virasoro subalgebra $V_c$. The actual stress tensor $T$ of the Kac-Moody algebra will be taken to be $T_\text{Sug}$, namely $T = T_\text{Sug}$. The eigenvalue under the action of $L_0$ will be denoted as $h$\footnote{Not to be confused with the Cartan generators $h^i$ and $h^i_n$.}, called the conformal weight. In subsequent discussions, we often omit the summation symbol $\sum$ over Lie algebra indices $a, b, \cdots$.

An affine primary $|\lambda\rangle$ is a state annihilated by $J_{n > 0}^A$ and $E^{\alpha > 0}_0$, with eigenvalues
\begin{align}
	h^i_{0} |\lambda\rangle = \lambda_i |\lambda \rangle, \qquad
	L_0 |\lambda\rangle = \frac{(\lambda, \lambda + 2 \rho)}{2(k + h^\vee)}|\lambda \rangle \ .
\end{align}
Starting from $|\lambda\rangle$ one can build a highest weight representation $M_\lambda$ of $\widehat{\mathfrak{g}}_k$ by acting on $|\lambda\rangle$ with $J^a_{n < 0}$ and $E^{\alpha < 0}_0$. In particular, the vacuum state $|0\rangle$ satisfies $J^a_{n \ge 0}|0\rangle = 0$, $L_{n \ge -1}|0\rangle = 0$, and it is the highest weight state of the vacuum representation $M_0$. Given a highest weight representation $M$, one can define its flavored character\footnote{The factor $y^k$ is an overall constant that encodes the level $k$, but is often omitted in physics literature, and also in many subsequent discussions in this paper. However, we keep in mind the existence of $y$ when defining modular transformations on the flavored characters.}
\begin{align}
	\operatorname{ch}_M(q, b)  \coloneqq y^k \operatorname{tr}_{M} q^{L_0 - c/24} b^{h} \ , \qquad b^h \coloneqq \prod_{i = 1}^{r} b_i^{h^i_0} \ .
\end{align}
Defined in this way the character $\operatorname{ch}_M(q, b)$ is a series in $q$ with certain region of convergence. Sometimes one can rewrite the character as the series expansion of an analytic function on the $q$-plane: different module characters may correspond to different expansions of the same analytic functions in different regions. Hence at the level of analytic functions, there is no one-to-one correspondence between modules and characters. Still, it is useful to study such analytic functional representation of characters if possible, which will be one of the main object of study in the following discussions. In general cases, the characters can be written as some combinations of quasi-Jacobi forms.

States in a highest weight module $M_\lambda$ besides $|\lambda\rangle$ are call affine descendants of $|\lambda\rangle$. If in $M_\lambda$ one finds a descendant state $|\mathcal{N}\rangle$ that also satisfies the affine primary condition, then $|\mathcal{N}\rangle$ will be called a singular vector, which implies the reducibility of $M_\lambda$ since $|\mathcal{N}\rangle$ serves as the highest weight state of a sub-representation. Descendants of a singular vector will be call null states. In general, any highest weight representation of $\widehat{\mathfrak{g}}_k$ also decomposes into irreducible representations of the Virasoro subalgebra $V_c$. In this sense, one can also define Virasoro singular vectors that are annihilated by $L_{n > 0}$, and the Virasoro null states as the Virasoro descendants.

By the state-operator correspondence, each state $|a\rangle$ in the vacuum module of $\widehat{\mathfrak{g}}_k$ corresponds to a local operator $a(z)$, where
\begin{align}
	a(z) = \sum_{n \in \mathbb{Z} - h_a} a_n z^{-n - h_a}, \quad a(z \to 0)|0\rangle = |a\rangle \ , \quad
	o(a) \coloneqq a_0 \ .
\end{align}
In particular, a null state $|\mathcal{N}\rangle$ in the vacuum module of $\widehat{\mathfrak{g}}_k$ corresponds to a null field $\mathcal{N}(z)$.

In the space of states, there is an interesting subspace $C_2(\widehat{\mathfrak{g}}_k)$, defined to be
\begin{equation}
  C_2(\widehat{\mathfrak{g}}_k) \coloneqq \operatorname{span}\{a_{ - h_a - 1}|b\rangle, a, b \in \widehat{\mathfrak{g}}_k\} \ .
\end{equation}
In terms of local operators, states in $C_2$ are just local normal-ordered composite operators with some derivative. It is a two-sided ideal with respect to the normal-ordered product \cite{zhu1996modular}. The quotient algebra $\widehat{\mathfrak{g}}_k/C_2$ is known as Zhu's $C_2$-algebra $R_{\widehat{\mathfrak{g}}_k}$ of $\widehat{\mathfrak{g}}_k$.

Null states are of great importance in classifying 2d CFTs. For example, when a null state is inserted into a plane correlation function, it turns into a differential equation that strongly constrains the correlation function involving only primaries \cite{Belavin:1984vu}. In this paper we will mainly encounter torus partition functions and one-point functions of Kac-Moody algebras, where null states also play a crucial role.

The torus one-point function of a null field (and its descendants) vanishes. Such a one-point function can sometimes be pre-processed by Zhu's recursion formula \cite{zhu1996modular,Mason:2008zzb,Beem:203X,Pan:2021ulr}. For a chiral algebra with a $U(1)$ affine current $J$, and any two operators $a(z), b(z)$ in the chiral algebra such that $J_0 |a \rangle = Q|a\rangle$,
\begin{align}
  \label{recursion2}
  \operatorname{str}_M o(a_{[- h_a]}b)x^{J_0}q^{L_0}
  = & \ \delta_{Q, 0}\operatorname{str}_M o(a) o(b) x^{J_0} q^{L_0} x^Q \nonumber\\
  & \ + \sum_{n = 1}^{+\infty} E_n\begin{bmatrix}
    e^{2\pi i h_a}\\x^Q
  \end{bmatrix} \operatorname{str}_M o(a_{[- h_a+n]}b)x^{J_0}q^{L_0} \ ,
\end{align}
and, when $n > 0$,
\begin{align}
	\operatorname{str}_M o(a_{[-h_a - n]} |b\rangle)q^{L_0} x^h
	= & \ (-1)^n \sum_{k = 1}^{+\infty} \begin{pmatrix}
	  k - 1\\n  
	\end{pmatrix}E_k \begin{bmatrix}
	  e^{2\pi i h_a}\\x^Q
	\end{bmatrix}\operatorname{str}_Mo(a_{[-h_a - n + k]}| b\rangle)q^{L_0}x^h \ . \nonumber
\end{align}
Here we have chosen a particular module $M$ of the chiral algebra to perform the (super)trace $\operatorname{str}$\footnote{For Kac-Moody algebra, $\operatorname{str}$ is simply the usual trace $\operatorname{tr}$. When fermionic generators are presence, states will be weighted by $(-1)^F$ based on the fermion number $F$.}. The square-mode $a_{[n]}$ follows from
\begin{align}
	a[z] \coloneqq e^{iz h_a} a(e^{iz} - 1) = \sum_{n \in \mathbb{Z} - h_a} a_{[n]} z^{-n - h_a} \ .
\end{align}
For a Virasoro primary $a$, the square modes $a_{[n]}$ have the same commutation relations among themselves as the standard mode $a_n$. In particular, $a_{[n]}|0\rangle = 0$ when $n > - h_a$. Also note that $o(L_{[-1]}a) = 0$ \cite{Gaberdiel:2008pr}. 

Some operator insertions can be equivalently rewritten as a differential operator acting on the character of the chosen module. For example,
\begin{align}
	\operatorname{str}o(L_{[-2]}^k |0\rangle) q^{L_0 - \frac{c}{24}} x^{J_0}
	= \mathcal{P}_k \operatorname{str} q^{L_0 - \frac{c}{24}}x^{J_0} \ ,
\end{align}
where
\begin{align}
	\mathcal{P}_1 = & \ q \partial_q = D_q^{(1)}, \qquad
	\mathcal{P}_2 = D_q^{(2)} + \frac{c}{2}E_4(\tau),\\
	\mathcal{P}_3 = & \ D_q^{(3)} + \left({8 + \frac{3c}{2}}\right)E_4(\tau)D_q^{(1)}
	+ 10c E_6(\tau) \ ,\\
  \mathcal{P}_4 = & \ D_q^{(4)} + (32 + 3c)E_4(\tau)D_q^{(2)} + (160 + 40c)E_6 D_q^{(1)} + (108c + \frac{3}{4}c^2)E_4(\tau)^2 \ .
\end{align}
Similarly, since $o(J_{[-1]}|0\rangle) = J_0$,
\begin{align}
	\operatorname{str}o(J_{[-1]} |0\rangle) q^{L_0 - \frac{c}{24}} x^{J_0}
	= D_x \operatorname{str} q^{L_0 - \frac{c}{24}}x^{J_0} \ , \qquad D_x \coloneqq x \partial_x \ ,
\end{align}

\subsection{Modular differential equations}

Differential equations have been an invaluable tool in classifying two dimensional conformal field theories. One well-known approach to classifying conformal blocks is the BPZ equation \cite{Belavin:1984vu}, which considers insertion of null field in plane correlation functions and turns it into a differential equations of primary correlators.

A similar approach to classifying RCFTs and their spectra is the holomorphic modular bootstrap \cite{Mathur:1988rx,Mathur:1988na,Eguchi:1987qd,Chandra:2018pjq,Mukhi:2020gnj,Das:2022bxm,Das:2021uvd,Das:2020wsi,Bae:2020xzl,Bae:2021mej,Duan:2022kxr,Duan:2022ltz,Duan:2022ltz,Gowdigere:2023xnm,Mukhi:2022bte,Das:2023qns}. The approach makes use of the modular invariance and the finite spectrum of a RCFT, which demands the characters of the chiral algebra to take value in a vector-valued modular form with respect to a suitable subgroup of $SL(2, \mathbb{Z})$. Using the Wronskian method, it then follows that any unflavored character $\operatorname{ch}$ must obey an $N$-th order ordinary differential equation, called an unflavored modular linear differential equation (MLDE),
\begin{align}
	(D_q^{(N)} + \sum_{r = 0}^{N - 1}\phi_{2(N - r)}(\tau) D_q^{(r)})\operatorname{ch} = 0 \ .
\end{align}
Here $\phi_k(\tau)$ denotes a weight-$k$ modular form with respect to a suitable modular group $\Gamma \subset SL(2, \mathbb{Z})$. Clearly, the unflavored MLDE is covariant under $\Gamma$, and it ties back to the fact that solutions of the equation forms a vector-valued modular form with respect to $\Gamma$.

Since the space of modular form at a given modular weight (and given pole structure in $\tau$, which is characterized by the Wronskian index) is finite dimensional with known basis, one may start by postulating a unflavored MLDE of certain weight, and work out all its solutions. With some additional physical constraints, one may find when the solutions to an unflavored MLDE give rise to the potential spectrum of a physically sensible conformal theory. By working one's way up in both modular weight and Wronskian index, a (partial) classification of the spectra of RCFTs can be obtained.

Non-rational CFTs, or more generally, non-rational chiral algebras exist. Moreover, non-rational chiral algebras arise prominently in the study of four dimensional $\mathcal{N} = 2$ superconformal field theories (SCFTs) as associated chiral algebras. Any 4d $\mathcal{N} = 2$ SCFT $\mathcal{T}$ always contains a nontrivial protected subsector of local operators which can be endowed with the structure of chiral algebra (or, vertex operator algebra) $\mathbb{V}[\mathcal{T}]$ under operator product expansion (OPE) restricted on a plane \cite{Beem:2013sza}. The chiral algebra $\mathbb{V}[\mathcal{T}]$ contains a Virasoro subalgebra with a negative central charge $c < 0$ and therefore the algebra is non-unitary, assuming the 4d SCFT is unitary. Further more, when the 4d theory has a non-trivial Higgs branch, the associated chiral algebra $\mathbb{V}[\mathcal{T}]$ will have an associated variety of positive dimension \cite{Beem:2017ooy}. The 4d Schur index $\mathcal{I}[\mathcal{T}]$ gives a weighted count of the operators in the subsector, and is identified with the vacuum character $\operatorname{ch}_0$. Based on these general properties and the many known examples \cite{Beem:2013sza,Lemos:2014lua,Beem:2014rza,Song:2017oew,Xie:2016evu,Cordova:2017mhb,Creutzig:2017qyf,Xie:2019zlb,Xie:2019vzr,Xie:2019yds,Kiyoshige:2020uqz}, it is believed that in general the associated chiral algebras are quasi-lisse \cite{Arakawa:2016hkg}, a significant generalization of the well-known rational chiral algebras.

A quasi-lisse chiral algebra $\mathbb{V}[\mathcal{T}]$ may have various type of modules. Some modules have simple structure that closely resemble those in a RCFT. For example, $L_0$ is diagonalizable and at any given eigenvalue $L_0 = h$ the subspace is finite dimensional. Such modules are referred to as ordinary modules. One can include flavor fugacities $b_i$ in the module character to keep track of the flavor charges of the states in the module, and for ordinary modules the character remains finite when $b_i$ are sent to one. It is known that the unflavored characters of ordinary modules of a quasi-lisse chiral algebra satisfy an unflavored MLDE \cite{Arakawa:2016hkg}, 
\begin{align}
	\Big(D_q^{(N)} + \sum_{\ell = 0}^{N - 1}\phi_{2(N - \ell)}(\tau) D_q^{(\ell)} \Big)\operatorname{ch} = 0 \ .
\end{align}
In many cases\footnote{There are scenarios where the null state $|\mathcal{N}_T\rangle$ does not lead to a differential equation, due to the presence of additional generators in $|\mathcal{N}_T\rangle$. One needs to go up in conformal weight to find another null state.}, via Zhu's recursion, such an equation arises from a null state $|\mathcal{N}_T\rangle$ in $\mathbb{V}[\mathcal{T}]$ that implements nilpotency of the stress tensor \cite{Beem:2017ooy},
\begin{equation}
	|\mathcal{N}_T\rangle = L_{-2}^N |0\rangle + c_2, \qquad c_2 \in C_2(\mathbb{V}[\mathcal{T}])\ .
\end{equation}
The equation has been used in classifying low rank 4d $\mathcal{N} = 2$ SCFTs \cite{Kaidi:2022sng}.

There are potentially non-ordinary modules, where the unflavoring limit $b_i \to 1$ of the flavored characters does not exist. In such a case, one has to keep flavor refinement. It is natural to expect that flavored characters in general are solutions to certain partial differential equations that generalize unflavored MLDEs. Indeed, evidences and examples have been worked out \cite{Zheng:2022zkm,Pan:2023jjw}, and among them are the non-unitary Kac-Moody algebras in the Deligne-Cvitanovi\'c exceptional series with level $k = - \frac{h^\vee}{6} - 1$. In these examples, the relevant equations contain (product of) Eisenstein series as their coefficients, which are of course quasi-Jacobi. We continue to call these equations flavored modular linear differential equations, as they are closely tied to the modular property of the characters. Having quasi-Jacobi form as coefficients leads to intricate behavior of the equations under a suitable modular group $\Gamma$. If we believe the characters of the associated chiral algebra enjoy certain modularity, then it must be encoded in the structure of the equations. In \cite{Pan:2023jjw,Zheng:2022zkm}, quasi-modular property of the equations has been observed in examples. Schematically, a flavored MLDE transforms into a sum of flavored MLDEs of equal or lower modular weight,
\begin{align}
	\operatorname{eq}_w \xrightarrow{S}& \ \operatorname{eq}_w + \operatorname{eq}_{w - 1} + \operatorname{eq}_{w - 2} + \cdots, \\
	\operatorname{eq}_{w - 1} \xrightarrow{S}& \ \operatorname{eq}_{w - 1} + \operatorname{eq}_{w - 2} + \cdots \ , \\
	& \ \vdots \nonumber
\end{align}
Quasi-modularity requires all of the equations on the right to annihilate all the flavored characters. They arise from additional null states of lower conformal weight. Together, the set of independent equations in the entire orbit places very strong constraint on the flavored characters.


\section{Holomorphic quasi-modular bootstrap}\label{section:examples}

Needless to say, null states play a crucial role in analyzing the flavored MLDEs of $\mathbb{V}[\mathcal{T}]$ or more generally quasi-lisse chiral algebras. Among all the null states, $|\mathcal{N}_T\rangle$ seems to feature most prominently, since it is guaranteed to exist (although it does not always give rise to a nontrivial equation), and in some cases it has been used to recover the ordinary spectrum of the chiral algebra.

Despite its importance, not much is known about the state $|\mathcal{N}_T\rangle$. It is defined by the equation
\begin{equation}\label{eq:nilpotency}
  |\mathcal{N}_T\rangle = L_{-2}^N |0\rangle + |c_2\rangle \ , \qquad c_2 \in C_2(\mathbb{V}[\mathcal{T}])  \ ,
\end{equation}
plus the requirement that $|\mathcal{N}_T\rangle$ is null. However, neither $N$ nor $c_2$ is known a priori: $N, c_2$ and $|\mathcal{N}_T\rangle$ has to be determined simultaneously such that $|\mathcal{N}_T\rangle$ is null. The goal of the following discussion is to propose a more intrinsic properties of $|\mathcal{N}_T\rangle$ for a class of algebras, which bootstrap the algebras and the spectra via flavored differential equations and quasi-modularity.

In \cite{Pan:2023jjw}, a simple defining property of $|\mathcal{N}_T\rangle$ is proposed when $|\mathcal{N}_T\rangle$ is a weight-four state in an affine Lie algebra $\widehat{\mathfrak{g}}_k$ associated to a simple Lie algebra $\mathfrak{g}$ \footnote{In this paper we focus on algebras associated to simple Lie algebras. It would be very interesting to explore if the intermediate affine Lie algebras \cite{Mathur:1988na,Chandra:2018pjq,Kawasetsu2014,Lee:2023owa,Lee:2024fxa,Sun:2024mfz} can be studied in a similar way described in this paper.},
\begin{equation}\label{eq:NT-conditions-weak}
  J^a_{0}|\mathcal{N}_T\rangle = L_{n \ge 1}|\mathcal{N}_T\rangle = 0 \ .
\end{equation}
Namely, $|\mathcal{N}_T\rangle$ is a $\mathfrak{g}$-singlet and a Virasoro primary with respect to the Sugawara stress tensor. The singlet condition is obvious since $L_{-2}^N|0\rangle$ is also a singlet. The above property allows one to determine simultaneously the state $|\mathcal{N}_T\rangle$, the allowed $\mathfrak{g}$, level $k$, the $S$-orbit of the $|\mathcal{N}_T\rangle$ equation and the entire spectrum. Unfortunately, the Virasoro primary condition cannot account for the $|\mathcal{N}_T\rangle$ in more general Kac-Moody algebra, since the solution to the condition is far from unique in general. Therefore, in this paper we would like to propose a stronger condition for $|\mathcal{N}_T\rangle$ in some class of Kac-Moody algebras. The conditions of $|\mathcal{N}_T\rangle$ we will be studying are
\begin{equation}\label{eq:NT-conditions}
  J^a_0|\mathcal{N}_T\rangle = L_2 |\mathcal{N}_T\rangle = h^i_{n \ge 2}|\mathcal{N}_T\rangle\ , \qquad i = 1, 2,..., r\ .
\end{equation}
Although this condition looks unrelated to equation (\ref{eq:nilpotency}), it turns out that in the cases we study, it uniquely determines the level $k$ and the central charge $c$, the form of $|\mathcal{N}_T\rangle$, guarantees $|\mathcal{N}_T\rangle$ to be a descendant of a singular vector, and it solves (\ref{eq:nilpotency}). When possible, we provide an elegant formula that relate $|\mathcal{N}_T\rangle$ with a singular vector of the algebra. We remark that the condition (\ref{eq:NT-conditions}) implies the weaker one (\ref{eq:NT-conditions-weak}). First of all, $h^i_{n \ge 2}|\mathcal{N}_T\rangle$ and the $\mathfrak{g}$-invariance implies $J^a_{n \ge 2}|\mathcal{N}_T\rangle = 0$. Therefore,
\begin{align}
  L_1 |\mathcal{N}_T\rangle = & \ \frac{1}{2(k + h^\vee)} K_{ab} \biggl(
    \sum_{m \ge 1} J_m^a J_{1 + m}^b + \sum_{m \ge 0}J^b_{1 - m}J^a_m
  \biggr)|\mathcal{N}_T\rangle \nonumber \\
  = & \ \frac{1}{2(k + h^\vee)} K_{ab}J^b_0J^a_1 |\mathcal{N}_T = \frac{1}{2(k + h^\vee)} K_{ab}f^{ba}{_c} J^c_1|\mathcal{N}_T\rangle = 0 \ .
\end{align}
Similarly,
\begin{align}
  L_{n \ge 3}|\mathcal{N}\rangle = & \ \frac{1}{2(k + h^\vee)} K_{ab} \left(
  \sum_{m \ge 1} J^a_{-m} J^b_{n + m}
  + \sum_{m \ge 0} J^b_{n - m} J^a_{m}
  \right) |\mathcal{N}\rangle \nonumber\\
  = & \ \frac{1}{2(k + h^\vee)} K_{ab} J^b_{n - 1}J^a_1|\mathcal{N}\rangle
  = \frac{1}{2(k + h^\vee)} K_{ab} f^{ba}{_c}J^c_{n}|\mathcal{N}\rangle = 0 \ .
\end{align}
Together with the explicit constraint $L_2|\mathcal{N}_T\rangle$ in (\ref{eq:NT-conditions}), it implies (\ref{eq:NT-conditions-weak}).

Using Zhu's recursion, the flavored MLDE corresponding to $|\mathcal{N}_T\rangle$ is determined. Such an equation alone is too weak to control the spectrum of the algebra. Luckily, it is straightforward to work out the inhomogeneous transformation of the equation under the action of $SL(2, \mathbb{Z})$, generated by
\begin{align}
  S: & \ \tau \to - \frac{1}{\tau}, \qquad \mathfrak{b}_i \to \frac{\mathfrak{b}_i}{\tau}, \qquad \mathfrak{y} \to \mathfrak{y} - \frac{1}{\tau} \frac{1}{2} \sum_{i,j = 1}^{r}K^{ij}\mathfrak{b}_i \mathfrak{b}_j \ ,\\
  T: & \ \tau \to \tau + 1 \ ,
\end{align}
where $b_i = e^{2\pi i \mathfrak{b}_i}$ are the flavor fugacities, and $y = e^{2\pi i \mathfrak{y}}$ is the fugacity probing the level $k$. We then require all components in the transformation, and all further components in the modular transformation of each existing components to annihilate the flavored characters. We may call such a property of the characters ``quasi-modularity''. Solving this set of partial differential equation, we find the potential spectrum of the algebra; in the cases that we encounter, this set of equations are strong enough to fully determine the known spectra. As a generalization of the holomorphic modular bootstrap, we simply refer to this procedure as the ``holomorphic quasi-modular bootstrap'', emphasizing the quasi-Jacobi nature of the equations. In the algebras we studied, apart from a few exceptions, the $S$-orbit of $|\mathcal{N}_T\rangle$ is sufficient to determine the entire spectrum. It would be very interesting to explore under what conditions $|\mathcal{N}_T\rangle$ is able to determine the full spectrum via quasi-modularity.

Every component that appears in the modular orbit of $|\mathcal{N}_T\rangle$ is simply an additional flavored MLDE, which presumably arises from an additional null state $|\mathcal{N}\rangle$ besides $|\mathcal{N}_T\rangle$. We can reorganize the transformation of any one of the equation as the equivalent transformation between null states, schematically,
\begin{equation}
  S|\mathcal{N}\rangle = \sum_{\mathcal{N}'} a_{\mathcal{N}'} |\mathcal{N}'\rangle \ ,
\end{equation}
where the formal coefficients $a_{\mathcal{N}'}$ are functions of $\tau, \mathfrak{b}_i$, and are the same across all the algebras we have studied. In this way, we reveal a hidden and universal modular action on the null states.

Besides modular transformations, we will also investigate the equations under translation $\sigma$ of flavor fugacities
\begin{equation}
  \mathfrak{b}_i \xrightarrow{\sigma} \mathfrak{b}_i + n_i \tau, \qquad
  \mathfrak{y} \xrightarrow{\sigma} \mathfrak{y} + \sum_{i,j=1}^{r}K^{ij}n_j \mathfrak{b}_i
  + \sum_{i,j = 1}^{r} \frac{1}{2}K^{ij} n_i n_j \tau \ ,
\end{equation}
Here $n_i$ are rational numbers. Such translation is similar to the spectral flow operation. 

When $n_i$ satisfy certain integral conditions, the translation preserves all the flavored MLDEs, implying the operation maps a character to another. Thus it is an efficient way to generate a more complete character spectrum. Similar to modular transformation, the above translation on flavored MLDEs is inhomogeneous. Therefore, we can rewrite the translation operation as a transformation between null states,
\begin{equation}
  \sigma |\mathcal{N}\rangle = \sum_{\mathcal{N}'} b_{\mathcal{N}'} |\mathcal{N}'\rangle \ .
\end{equation}
Again, we will work out explicitly the coefficients $b_{\mathcal{N}'}$ and find that they are universal.

When $n_i$ satisfy certain half-integral conditions, the translation slightly changes the flavored MLDEs, by changing some signs of the twist parameter in the Eisenstein series. This operation potentially produces the twisted flavored MLDEs for twisted modules of the chiral algebra. Using the twisted equations, we determine the twisted spectrum of the algebra, gaining insight into the twisted sector.

\section{The Deligne-Cvitanovi\'{c} exceptional series}

Let us consider a null state $|\mathcal{N}_T\rangle$ in $\widehat{\mathfrak{g}}_k$ of weight-four. A similar analysis of this simplest case has been carried out in \cite{Pan:2023jjw}, here we will briefly go over the analysis but with the stronger conditions \eqref{eq:NT-conditions} proposed above. Since the state $|\mathcal{N}_T\rangle$ is expected to be $\mathfrak{g}$-invariant, we can write down its most general form\footnote{One may argue that in general there should be $K_{AC}K_{BD}$ and $K_{AD}K_{BC}$ terms. However, up to terms in $C_2(\widehat{\mathfrak{g}}_k)$, they are identical as the $K_{AB}K_{CD}$ term. Any $\mathfrak{g}$-invariant $C_2$ terms can be absorbed into the other terms already in $|\mathcal{N}_T\rangle$.}
\begin{align}
  |\mathcal{N}_T\rangle = (\beta K_{AB} J_{-3}^A J_{-1}^B & \ + \gamma K_{AB} J_{-2}^A J_{-2}^B + \delta K_{AB}K_{CD} J^A_{-1}J^B_{-1}J^C_{-1}J^D_{-1})|0\rangle \nonumber \\
  & \ + \delta_1 d_{ABC}J_{-2}^A J_{-1}^B J_{-1}^C|0\rangle + \delta_2 d_{ABCD}J_{-1}^A J_{-1}^B J_{-1}^C J_{-1}^D|0\rangle \ . \nonumber
\end{align}
For non-critical Kac-Moody algebras, we always impose the Sugawara construction,
\begin{equation}
  0 = L_{-2}|0\rangle - \frac{1}{2(k + h^\vee)}\sum_{A,B} K_{AB} J^A_{-1}J^B_{-1}|0\rangle \ ,
\end{equation}
equating the stress tensor $T$ with the Sugawara stress tensor
\begin{equation}
  T_\text{Sug} \coloneqq \frac{1}{2(k + h^\vee)}\sum_{A,B} K_{AB} (J^A_{-1}J^B_{-1}) \ .
\end{equation}

It is straightforward to show that for $ N \ge 1$,
\begin{equation}
  L_{-2}^N|0\rangle = \Bigg[\frac{1}{2(k + h^\vee)}\Bigg]^N (K_{AB}J_{-1}^AJ^B_{-1})^N|0\rangle + |c_2\rangle \ , \qquad |c_2\rangle \in C_2(\widehat{\mathfrak{g}}_k) \ .
\end{equation}
Hence we may replace the complicated $\gamma$ term with the simpler $L_{-2}^2 |0\rangle$ while $|c_2\rangle$ can be absorbed into the remaining $\alpha, \beta, \gamma, \delta_1$ terms. We have, after suitable normalization,
\begin{align}
  |\mathcal{N}_T\rangle = (L_{-2}^2 + & \ \beta K_{AB} J_{-3}^A J_{-1}^B +  \gamma K_{AB}J_{-2}^AJ_{-2}^B)|0\rangle \nonumber \\
  & \ + \delta_1 d_{ABC}J_{-2}^A J_{-1}^B J_{-1}^C|0\rangle + \delta_2 d_{ABCD}J_{-1}^A J_{-1}^B J_{-1}^C J_{-1}^D)|0\rangle\ .
\end{align}

Now we apply the conditions to constrain the coefficients $\alpha, \beta, \gamma, \delta_{1,2}$. First of all, the state is by construction $\mathfrak{g}$-invariant, and therefore $J_0^A |\mathcal{N}_T\rangle = 0$. We are left with
\begin{align}
  L_2|\mathcal{N}_T\rangle =& \ (8 + c + 6(k + h^\vee)\beta) L_{-2}|0\rangle\ ,
\end{align}
\begin{align}
  h_2^i |\mathcal{N}_T\rangle =  & \ (4k\gamma +(\beta + 2\gamma)h^\vee)J^i_{-2}|0\rangle \\
  & \ + 2k \delta_1 d_{ABC} K^{iA} J^B_{-1}J^C_{-1}|0\rangle \nonumber\\
  & \ + \delta_1 d_{ABC} f^{iA}{_{C_1}}f^{C_1 C}{_{C_2}} J^B_{-1} J^{C_2}_{-1}|0\rangle\nonumber\\
  & \ + \delta_1 d_{ABC} f^{iA}{_{C_1}}f^{C_1 B}{_{C_2}}J^{C_2}_{-1}J^C_{-1}|0\rangle\nonumber\\
  & \ + \delta_2 d_{ABCD}f^{i B}{_{C_1}}f^{C_1 C}{_{C_2}}f^{C_2 D}{_{C_3}}J^A_{-1}J^{C_3}_{-1}|0\rangle\nonumber\\
  & \ + \delta_2 d_{ABCD} f^{iA}{_{C_1}} f^{C_1 C}{_{C_2}}f^{C_2 D}{_{C_3}}J^B_{-1}J^{C_3}_{-1}|0\rangle\nonumber\\
  & \ + \delta_2 d_{ABCD} f^{iA}{_{C_1}}f^{C_1 B}{_{C_2}}f^{C_2 D}{_{C_3}}J^C_{-1}J^{C_3}_{-1}|0\rangle\nonumber\\ 
  & \ + \delta_2 d_{ABCD} f^{iA}{_{C_1}}f^{C_1 B}{_{C_2}}f^{C_2 C}{_{C_3}}J^{C_3}_{-1}J^D_{-1}|0\rangle \ ,
\end{align}
and
\begin{align}
  h_3^i |\mathcal{N}_T\rangle = & \ (3 + 3k\beta + 2 h^\vee(\beta + \gamma))J^i_{-1}|0\rangle \nonumber\\
  & \ + \delta_1 d_{ABC} f^{iA}{_{C_1}}f^{C_1B}{_{C_2}}f^{C_2 C}{_{C_4}}J^{C_4}_{-1}|0\rangle \nonumber\\
  & \ + \delta_2 d_{ABCD} f^{iA}{_{C_1}}f^{C_1B}{_{C_2}}f^{C_2 C}{_{C_3}}f^{C_3 D}{_{C_4}}J^{C_4}_{-1}|0\rangle \ ,\\
  h_4^i |\mathcal{N}_T\rangle = & \ 0  \  \ .
\end{align}
For all of these states to vanish, we need
\begin{align}
  0 = \delta_1 = \delta_2 = & \ 8 + c + 6(k + h^\vee)\beta
  = 4k \gamma + (\beta + 2\gamma)h^\vee \nonumber\\
  = & \ 3 + 3k \beta + 2 h^\vee (\beta + \gamma) \ .
\end{align}
With the Sugawara central charge $c = k\dim\mathfrak{g}/(k + h^\vee)$, we can solve
\begin{align}
  \beta = & \ - \frac{3(2k + h^\vee)}{(k + h^\vee)(6k + h^\vee)}, \qquad
  \gamma = \frac{3h^\vee}{2(k + h^\vee)(6k + h^\vee)}, \\
  \dim \mathfrak{g} = & \ - \frac{2(6k - 5h^\vee)(k + h^\vee)}{k(6k + h^\vee)} \ .
\end{align}
In particular,
\begin{equation}
  \beta - 2 \gamma = - \frac{6}{6k + h^\vee} \ .
\end{equation}

Now we can apply Zhu's recursion formula to $|\mathcal{N}_T\rangle$, and obtain a flavored modular differential equation,
\begin{align}\label{eq:NT-weight-4}
  0 = & \ D_q^{{(2)}} \operatorname{ch} + \Bigl(\frac{c}{2} + 3kr (\beta - 2 \gamma)\Bigr)E_4(\tau)\operatorname{ch} \\
  & \ + (\beta - 2\gamma) \sum_{\alpha \in \Delta}\sum_{i = 1}^{r}K_{\alpha, -\alpha} f^{\alpha, -\alpha}{_i}E_3 \begin{bmatrix}
    1 \\ b^\alpha
  \end{bmatrix} D_{b_i}\operatorname{ch} 
  + 3k (\beta - 2 \gamma)\sum_{\alpha \in \Delta} E_4 \begin{bmatrix}
    1 \\ b^\alpha
  \end{bmatrix} \operatorname{ch} \ . \nonumber
\end{align}
Additionally, the Sugawara construction $T - T_\text{Sug} = 0$ gives another equation,
\begin{align}\label{eq:Sugawara}
  0 = 2(k + h^\vee)D_q^{(1)} \operatorname{ch} - K_{ij} D_{b_i}D_{b_j} \operatorname{ch} - \sum_{\alpha}K_{\alpha, -\alpha} & f^{\alpha, -\alpha}{_i} E_1 \begin{bmatrix}
    1 \\ b^\alpha
  \end{bmatrix} D_{b_i}\operatorname{ch} \\
  & \ - kr E_2(\tau) \operatorname{ch} - k \sum_{\alpha}E_2 \begin{bmatrix}
    1 \\ b^\alpha
  \end{bmatrix} \operatorname{ch} \ . \nonumber
\end{align}

Quasi-modularity requires the character $\operatorname{ch}$ to satisfy the $S$-transformed equation \eqref{eq:NT-weight-4}, \eqref{eq:Sugawara}. In particular, (\ref{eq:NT-weight-4}) transforms into sum of expressions of weights from $0$ to $4$, and the character should satisfy each expression separately. The zero-weight part is particularly simple as it does not contain any derivative,
\begin{equation}
  (\mathfrak{b}^2)^2 + \frac{i(\beta - 2\gamma)}{48\pi^3}
  \sum_{\alpha} \sum_{i=1}^{r} K_{\alpha, - \alpha}f^{\alpha, -\alpha}{_i} (\ln b^\alpha)^3 (\partial_{\mathfrak{b}_i} \mathfrak{b}^2)
  - \frac{3k(\beta - 2\gamma)}{384\pi^4}  \sum_{\alpha} (\ln b^\alpha)^4\ . 
\end{equation}
where $\mathfrak{b} \coloneqq \sum_{i = 1}^{r}\mathfrak{b}_i\alpha_i^\vee$, $\mathfrak{b}^2 \coloneqq \frac{k}{2}(\mathfrak{b}, \mathfrak{b})$. For this to vanish it implies a surprisingly strong constraint on the root structure of the finite Lie algebra $\mathfrak{g}$, 
\begin{equation}
  \lambda_\mathfrak{g}  (\mathfrak{b}, \mathfrak{b})^2 = \sum_{\alpha}(\alpha, \mathfrak{b})^4 \ ,
\end{equation}
where $\lambda_\mathfrak{g}$ is some constant independent of $\mathfrak{b}$. Only the algebras from the Deligne-Cvitanovi\'{c} series \cite{Deligne} satisfy this constraint,
\begin{equation}
  \mathfrak{a}_1, \quad \mathfrak{a}_2, \quad \mathfrak{g}_2, \quad \mathfrak{d}_4, \quad
  \mathfrak{f}_4, \quad \mathfrak{e}_{6, 7, 8} \ ,
\end{equation}
with $\lambda_\mathfrak{g} = 6(\frac{h^\vee}{6} + 1)$. This further fixes $k$ and the coefficients in $|\mathcal{N}_T\rangle$ to be either
\begin{align}
  k = & \ +1, \qquad \dim \mathfrak{g} = \frac{2(h^\vee + 1)(5h^\vee - 6)}{h^\vee + 6} \ , \\
  \beta = & \ - \frac{3(h^\vee + 2)}{(6+h^\vee)(1 + h^\vee)}, \quad
  \gamma = \frac{3h^\vee}{2(6+h^\vee)(1 + h^\vee)} \ ,
\end{align}
or
\begin{align}
  k = & \ - \frac{h^\vee}{6} - 1, \qquad \dim \mathfrak{g} = \frac{2(h^\vee + 1)(5h^\vee - 6)}{h^\vee + 6} \ , \\
  \beta = & \ \frac{2(h^\vee - 3)}{5h^\vee - 6}, \quad
  \gamma = - \frac{3h^\vee}{2(5h^\vee - 6)} \ .
\end{align}
We note that the above gives precisely the dimension formula \cite{Cohen1996ComputationalEF}. Therefore analysis of the flavored MLDEs recovers the unitary and non-unitary Deligne-Cvitanovi\'{c} series of chiral algebras \cite{Mathur:1988na,Tuite:2006xi,Tuite:2014fha,Beem:2017ooy,Arakawa:2016hkg,Abhishek:2023wzp}. The weight-one part in the $S$-transformed (\ref{eq:NT-weight-4}) automatically vanishes as long as the above data are fixed. What remain are the weight-two and three, which provide genuine differential constraints on the characters,
\begin{align}\label{eq:weight-3-DC}
  0 = & \ \Bigg(2  \sum_i \mathfrak{b}_i D_{b_i}(D_q^{(1)} + E_2)
  -  (\beta - 2 \gamma) \sum_{\alpha, i} \frac{|\alpha_i|^2}{2} m_i^\alpha  E_2 \begin{bmatrix}
    1 \\ b^\alpha  
  \end{bmatrix}(\alpha, \mathfrak{b}) D_{b_i} \nonumber \\
  & \ \qquad \qquad\qquad \qquad\qquad\qquad\qquad - 2 k  (\beta - 2\gamma) \sum_{\alpha} (\alpha, \mathfrak{b}) E_3 \begin{bmatrix}
    1 \\ b^\alpha  
  \end{bmatrix}\Bigg) \operatorname{ch} \ ,
\end{align}
and
\begin{align}\label{eq:weight-2-DC}
  0 = \Bigg(k & (\mathfrak{b}, \mathfrak{b}) (D_q^{(1)} + E_2)+ \mathfrak{b}_i \mathfrak{b}_j D_{b_i} D_{b_j} + \frac{1}{2}(\beta - 2\gamma) \sum_{\alpha, i} \frac{|\alpha_i|^2}{2}m_i^\alpha (\alpha, \mathfrak{b})^2 E_1 \begin{bmatrix}
    1 \\ b^\alpha  
  \end{bmatrix}D_{b_i} \nonumber \\ 
  & \ + \frac{1}{2}k (\beta - 2 \gamma) \sum_{\alpha} (\alpha, \mathfrak{b})^2 E_2 \begin{bmatrix}
    1 \\ b^\alpha  
  \end{bmatrix}\Bigg)\operatorname{ch} \ .
\end{align}
Note that $\mathfrak{b}_i$ are arbitrary flavor fugacities, hence the above actually implies a set of $r$ and $\frac{r(r+1)}{2}$ independent partial differential equations. As noticed in \cite{Pan:2023jjw}, these equations correspond to additional null states in the chiral algebra. Using the corresponding null states, we may now write concretely the effective $S$-action on the null state $|\mathcal{N}_T\rangle$,
\begin{equation}
  S|\mathcal{N}_T\rangle = \sum_{\ell\ge 0} \sum_{i_1, ... i_\ell = 1}^{r}  \frac{1}{\ell!} \tau^{h[\mathcal{N}_T] - \ell} \mathfrak{b}_{i_1} \cdots \mathfrak{b}_{i_\ell} h^{i_1}_1 ... h^{i_\ell}_1 |\mathcal{N}_T\rangle \ .
\end{equation}
In fact, starting from any other null state, such as $h_1^i|\mathcal{N}_T\rangle$ or $h_1^ih_1^j|\mathcal{N}_T\rangle$, the $S$-action is the same,
\begin{equation}
  S|\mathcal{N}\rangle = \sum_{\ell\ge 0} \sum_{i_1, ... i_\ell = 1}^{r}  \frac{1}{\ell!} \tau^{h[\mathcal{N}] - \ell} \mathfrak{b}_{i_1} \cdots \mathfrak{b}_{i_\ell} h^{i_1}_1 ... h^{i_\ell}_1 |\mathcal{N}\rangle \ .
\end{equation}
One can easily check that $S^2 = 1$,
\begin{equation}
  S^2 |\mathcal{N}\rangle =
  \sum_{\substack{\ell \ge 0 \\ \ell' \ge 0}}
  \sum_{\substack{i_1 ... i_\ell = 1 \\ j_1 ... j_{\ell'} = 1}}^{r} \frac{(-1)^\ell}{\ell! \ell'!}\tau^{ - \ell - \ell'} \mathfrak{b}_{j_1}\dots \mathfrak{b}_{j_{\ell'}} \mathfrak{b}_{i_1}\dots\mathfrak{b}_{i_\ell} h_1^{j_1} ... h_1^{j_{\ell'}} h_1^{i_1}\dots h_1^{i_\ell}|\mathcal{N}\rangle = |\mathcal{N}\rangle \ ,
\end{equation}
using
\begin{equation}
  \sum_{\ell \ge 0}^{h_\mathcal{N}} \frac{(-1)^\ell}{\ell! (L - \ell)!} = \delta_{L,0} \ , \qquad L = 0, 1, 2, \dots \ .
\end{equation}
The effective $T$-action acts only on the fugacity $\tau \to \tau + 1$, and it is easy to check that $(ST)^3$ on the flavored MLDEs and null states gives $1$.

This set of flavored modular differential equations is highly constraining, as they are enough to completely determine all the module characters corresponding to the irreducible modules at the integrable level $k = 1$, the admissible characters for $\widehat{\mathfrak{su}}(2)_{-4/3}$ and $\widehat{\mathfrak{su}}(3)_{-3/2}$ \cite{Kac:1988qc,2016arXiv161207423K,Cordova:2017mhb}, as well as the characters of the modules discussed in \cite{AXTELL2011195,pervse2007vertex,pervse2013note,Arakawa:2015jya,Peelaers}. Some additional remarks follow.

\subsection{\texorpdfstring{$\widehat{\mathfrak{g}}_1$}{}}

Let us go into some detail. The level $k = 1$ Kac-Moody algebra $\widehat{\mathfrak{g}}_1$ are well-known rational conformal field theories. The vacuum modules all contain a singular vector at conformal weight-2,
\begin{equation}
  |v_\text{sing}\rangle = (E_{-1}^\theta)^{k + 1} |0\rangle \ , \qquad k = 1\ .
\end{equation}
Under the action of the finite algebra $\mathfrak{g}$, it generates a representation $\mathcal{R}_{2\theta}$, whose highest weight has the same Dynkin label as $2\theta$. The $S$-transformation of the $|\mathcal{N}_T\rangle$-equation generates those equations of $h_1^ih_1^j|\mathcal{N}_T\rangle$. The Sugawara equation is outside of the $S$-orbit of $|\mathcal{N}_T\rangle$, as no linear combination of $h_1^ih_1^j|\mathcal{N}_T\rangle$ reproduces $T - T_\text{Sug}$. This is tied to the fact that there exist the non-trivial singular vector $|v_\text{sing}\rangle$ and its descendants $v^{ab}$, which contains the non-zero states $h_1^ih_1^j|\mathcal{N}_T\rangle$ as some of the zero-charge states. As a result, there are additional flavored modular differential equations that are not accounted for by the $S$-orbit of $|\mathcal{N}_T\rangle$, including the Sugawara equation and those arising from some other zero-charge states in $\mathcal{R}_{2\theta}$.

Interestingly, although far from being the full set of equations from null states, the equations (\ref{eq:Sugawara}), (\ref{eq:NT-weight-4}), (\ref{eq:weight-3-DC}) and (\ref{eq:weight-2-DC}) together determine all the integrable module characters. Consider the standard ansatz
\begin{equation}
  \operatorname{ch} = q^\sigma \sum_{n \ge 0} c_n(b_1, \cdots, b_r)q^n\ , 
\end{equation}
To leading order, the $|\mathcal{N}_T\rangle$-equation (\ref{eq:NT-weight-4}) imposes
\begin{equation}
  c + 6(240\sigma^2 - 40\sigma + \dim \mathfrak{g}(\beta - 2\gamma)) = 0\ .
\end{equation}
Inserting the Sugawara central charge and the solutions to $\beta, \gamma$ for $k = 1$, there are two solutions for $\sigma$,
{\small\begin{equation}
  \sigma = \frac{2 h^\vee (h^\vee+7)+12 \pm \sqrt{2} \sqrt{(h^\vee+1) (h^\vee+6) (\dim \mathfrak{g} (7 h^\vee+6)+2 (h^\vee+1) (h^\vee+6))}}{24 (h^\vee+1) (h^\vee+6)} \nonumber \ .
\end{equation}}
Substituting the Lie algebra data for the Deligne-Cvitanovi\'{c} series, $\sigma + \frac{1}{24}c$ are precisely the conformal weight of the highest weight states of the integrable modules. Additionally, the $|\mathcal{N}_T\rangle$-equation can be used to algebraically solve $c_{n}(b)$ in terms of $c_0(b)$, $c_1(b)$, $ ...$, $c_{n - 1}(b)$ recursively, since at the $n$-th order, $c_n(b)$ appears always without derivative with respect to $b_i$.  Hence we only need to solve the leading coefficient $c_0(b)$.

The two weight-two equations (\ref{eq:Sugawara}) and (\ref{eq:weight-2-DC}) impose conditions on the $n_i$ in the ansatz for $c_0$ as $a_i$-series,
\begin{equation}
  c_0(b_1,\dots, b_r) = a_1^{n_1} \dots a_r^{n_r} \sum_{\ell_i\ge 0} c_{0;\ell_1, \dots, \ell_r} a_1^{\ell_1} \dots a_r^{\ell_r} \ ,
\end{equation}
respectively,
\begin{align}
  0 = & \ 2(k + h^\vee) \sigma - \sum_{i,j}(\mathbf{n},\mathbf{n})
  + \sum_{\alpha > 0} (\alpha, \mathbf{n}) + \frac{k}{12} \dim \mathfrak{g}\ , \\
  0 = & \ k\left[\sigma - \frac{1}{12} - \frac{h^\vee}{12}(\beta - 2\gamma)\right] (\mathfrak{b}, \mathfrak{b})
  - \frac{1}{2}(\beta - 2\gamma) \sum_{\alpha > 0}(\mathbf{n}, \alpha)(\alpha, \mathfrak{b})^2
  + (\mathfrak{b}, \mathbf{n})^2 \ , 
\end{align}
where $\mathbf{n} \coloneqq \sum_{j}n_j \alpha_j$, and the $a_i$ fugacities are related to the $b_i$ by
\begin{align}
  a_i \coloneqq \prod_{j} b_j^{(\alpha_i, \alpha_j^\vee)} \ , \qquad
  D_{b_j} = \sum_{n = 1}^{r} \frac{|\alpha_n|^2}{2} K^{nj} D_{a_n} \ .
\end{align}
The solution of $\mathbf{n}$ to these two equations precisely reproduces the highest weights of the integrable modules of the Deligne-Cvitanovi\'{c} Kac-Moody algebras.

\subsection{\texorpdfstring{$\widehat{\mathfrak{g}}_{k = - \frac{h^\vee}{6} - 1}$}{}}

The non-unitary Deligne-Cvitanovi\'{c} series have close relation with 4d $\mathcal{N} = 2$ physics. Except for the non-simply laced entries, $\widehat{\mathfrak{g}}_{- \frac{h^\vee}{6} - 1}$ is the associated chiral algebra of a 4d $\mathcal{N} = 2$ SCFT, whose associated variety, being the minimal nilpotent orbits of $\mathfrak{g}$, is identified with the Higgs branch \cite{Beem:2017ooy}. The series contains two boundary admissible algebras $\widehat{\mathfrak{su}}(2)_{-4/3}, \widehat{\mathfrak{su}}(3)_{-3/2}$, while the rest are non-admissible with negative integral level. The non-unitary Deligne-Cvitanovi\'{c} series share many similar features as the $k = 1$ counterparts. Apart from $\widehat{\mathfrak{su}}(2)_{-4/3}$ having a weight-three singular vector, all the other algebras in the series have their nontrivial singular vectors $v_\text{sing}$ at conformal weight $h_\text{sing} = 2$. They are known as the Joseph ideal \cite{ASENS_1976_4_9_1_1_0,Arakawa:2015jya}
\begin{equation}
  (J^{a}J^{b})\Big|_{\mathfrak{R}} = 0 \ ,
\end{equation}
where the representation $\mathfrak{R}$ follows from the decomposition $\operatorname{symm}^2\mathbf{adj} = \mathbf{1} \oplus \mathcal{R}_{2\theta} \oplus \mathfrak{R}$. The null states $h_1^ih_1^j|\mathcal{N}_T\rangle$ at $h = 2$ are inside $\mathfrak{R}$, accounting for all the charge-zero states in $\mathfrak{R}$. For the ADE subseries, the Sugawara equation is actually in the $S$-orbit of $|\mathcal{N}_T\rangle$, but not for $\mathfrak{g} = \mathfrak{g}_2, \mathfrak{f}_4$. This can be seen from the simple fact that
\begin{align}
  \# \text{zero-charge in $\mathfrak{R}_\text{ADE}$} + 1 = & \ \frac{r(r+1)}{2} \ , \\
  \# \text{zero-charge in $\mathfrak{R}_{\mathfrak{g}_2, \mathfrak{f}_4}$} + 1 > & \ \frac{r(r+1)}{2} \ .
\end{align}
In other words, for the ADE cases, the $\frac{r(r+1)}{2}$ states $h^i_1h^j_1|\mathcal{N}_T\rangle$ account for the states of zero charges in $\mathfrak{R}_\text{ADE}$ and additionally the trivial Sugawara condition $T - T_\text{Sug}$. But for non-ADE cases, those $\frac{r(r+1)}{2}$ states can only take care of those in $\mathfrak{R}_{\mathfrak{g}_2, \mathfrak{f}_4}$.

Since the Sugawara equation falls within the $S$-orbit of $|\mathcal{N}_T\rangle$, the spectrum of the $ADE$ subseries is completely determined by $|\mathcal{N}_T\rangle$ alone with quasi-modularity. For the $\mathfrak{g}_2, \mathfrak{f}_4$, the $S$-orbit of $|\mathcal{N}_T\rangle$ together with the Sugawara equation completely determine the spectrum.

\section{\texorpdfstring{$\widehat{\mathfrak{su}}(2)_{k}$ algebras}{}}

\subsection{General properties}
It is straightforward to go beyond the conformal-weight-four $|\mathcal{N}_T\rangle$, however, it is a computationally heavy task in full generality. For simplicity, in this paper we focus on the simplest Kac-Moody algebras $\widehat{\mathfrak{su}}(2)_k$ with $k$ to be determined, and leave the more complete classification to future work. We consider the state $|\mathcal{N}_T\rangle$ of conformal weight $2n_T$, and require the following set of $|\mathcal{N}_T\rangle$-conditions,
\begin{equation}
  0 = J^a_0|\mathcal{N}_T\rangle = L_2 |\mathcal{N}_T\rangle = h^{i = 1}_{n \ge 2}|\mathcal{N}_T\rangle\ .
\end{equation}
For $SU(2)$, we take indices $a,b, \ldots = 1,2,3$, or more often $+, -, 1$. Concretely, $J_n^\pm \coloneqq J_n^1 \pm i J_n^2$, $h^{i = 1}_n \coloneqq 2 J^{a = 3}_n$; here we use $h^i_n$ to avoid confusing $J^{i = 1}_n$ with $J^{a = 1}_n$.

We observe from explicit computation that these simple conditions impose strong constraints on both $|\mathcal{N}_T\rangle$ and the level $k$. The exact expression of $|\mathcal{N}_T\rangle$ is completely fixed and the allowed values $k$ are simultaneously determined. Here we tabulate the results with simplest values of $k$.
\begin{table}[h]
  \centering
  \begin{tabular}{c|l}
    $h[\mathcal{N}_T]$ & allowed $k$\\
    \hline
    4 & $1$, $- 4/3$\\
    6 & $2$, $- 8/5$, $- 1/2$\\
    8 & $3$, $-12/7$\\
    10 & $4$, $-16/9$, $- 5/4$\\
    12 & $5$, $- 20/11$, $-7/5$, $-2/3$, $1/2$
  \end{tabular}
\end{table}
From the table, we recover known Kac-Moody algebras with integrable level $k = 1, 2, ...$ and fractional admissible levels $k = \frac{t}{v}$, including both boundary (such that $k + h^\vee = \frac{h^\vee}{v}$) and non-boundary admissible. Based on the computational data, we summarize crucial properties of $|\mathcal{N}_T\rangle$ for these algebras in the following.
\begin{itemize}
  \item For the integral levels $k$, it is well known that the singular vector is given by $v_\text{sing} = (J_{-1}^+)^{k + 1}|0\rangle$. Under the action of the finite $\mathfrak{su}(2)$ subalgebra, it generates a spin-$(k+1)$ representation, whose basis can be denoted as $|v^{a_1 ... a_{k + 1}}\rangle$ which is fully symmetrized and traceless in the indices \footnote{Explicitly, take $k = 1$ as an example, we can write $v^{++} = v_\text{sing}$, and
  \begin{align}
    |v^{+ 1}\rangle = & \ |v^{1 +}\rangle = - \frac{1}{2} J_0^-|v^{++}\rangle \ , \qquad
    &|v^{11}\rangle = & \frac{1}{3} J_0^-J_0^-|v^{++}\rangle \ ,\\
    |v^{1-}\rangle = & \ \frac{1}{12} J_0^-J_0^-J_0^-|v^{++}\rangle \ , 
    &|v^{--}\rangle = & \frac{1}{24} J_0^-J_0^-J_0^-J_0^-|v^{++}\rangle \ .
  \end{align}
  }. Then the null state $|\mathcal{N}_T\rangle$ is given by
  \begin{equation}
    |\mathcal{N}_T\rangle = K_{a_1 b_1} ... K_{a_{k + 1} b_{k+1}} J_{-1}^{b_1} ... J_{-1}^{b_{k + 1}} |v^{a_1 ... a_{k + 1}}\rangle \ .
  \end{equation}

  \item The boundary admissible levels are $k = - 4n/(2n + 1)$, $n = 1, 2, ...$. The singular vector $|v_\text{sing}\rangle$ in a boundary admissible $\widehat{\mathfrak{su}}(2)_k$ generates a spin-$1$ representation $\mathbf{3}$, whose basis can be written as $|v^a\rangle$. The null state $|\mathcal{N}_T\rangle$ is given by
  \begin{equation}
    |\mathcal{N}_T\rangle = K_{ab} J_{-1}^a |v^b\rangle \ .
  \end{equation}
  \item For general admissible $\widehat{\mathfrak{su}}(2)_k$, the singular vector belongs to a higher-spin representation of the finite $\mathfrak{su}(2)$, whose basis vectors can be written as $|v^{a_1 ... a_{\ell}}\rangle$. The null state $|\mathcal{N}_T\rangle$ is again given by
  \begin{equation}
    |\mathcal{N}_T\rangle = K_{a_1 b_1} ... K_{a_{\ell} b_{\ell}} J_{-1}^{a_1} ... J_{-1}^{a_{\ell}} | v^{b_1 ... b_{\ell}}\rangle \ .
  \end{equation}
\end{itemize}

Let us go into some details. For the integrable $\widehat{\mathfrak{su}}(2)_{k = 1, 2, 3, ...}$, the states $|v^{a_1 ... a_{k + 1}}\rangle$ can be explicitly written as
  \begin{equation}
    J^{(a_1}_{-1} ... J^{a_{k + 1})}_{-1}|0\rangle - \text{traces} \ ,
  \end{equation}
  hence $|\mathcal{N}_T\rangle$ is simply
  \begin{equation}
    |\mathcal{N}_T\rangle = \sum_{\sigma \in S_{k + 1}} \Lambda_\sigma K_{a_1 b_{\sigma(1)}} ... K_{a_{k + 1} b_{\sigma(k + 1)}} J_{-1}^{b_{\sigma(1)}}  ... J_{-1}^{b_{\sigma(k + 1)}} J^{(a_1}_{-1} ... J^{a_{k + 1})}_{-1}|0\rangle \ ,
  \end{equation}
  with some rational coefficients $\Lambda_\sigma$ depending on the permutation $\sigma \in S_{k + 1}$ of the $k + 1$ indices. Using commutation relations, this sum can be reorganized into the form
  \begin{align}
    |\mathcal{N}_T\rangle
    = & \ \left({\frac{1}{2(k + h^\vee)}}\right)^{n + 1} K_{a_1 b_1} ... K_{a_{k + 1} b_{k + 1}} J^{b_1}_{-1} ... J^{b_{k + 1}}_{-1} J^{a_1}_{-1} ... J^{a_{k + 1}}_{-1} |0\rangle + |c_2\rangle \nonumber \\
    = & \ L_{-2}^{n + 1} |0\rangle + |c'_2\rangle \ ,
  \end{align}
  implementing the nilpotency.

For the admissible $\widehat{\mathfrak{su}}(2)_k$ algebras, some properties of the singular vectors $|v_\text{sing}\rangle$ of are known from math literature \cite{KAC197997,Malikov}. Concretely, define $k + 2 = \frac{u}{v}$ for coprime integers $u, v$; for admissible levels, $u \ge h^\vee = 2$. The singular vector $|v_\text{sing}\rangle$ in the vacuum module has conformal weight and finite $SU(2)$ weight (Dynkin label) given by
\begin{equation}
  h_\text{sing} = (u-1)v, \qquad \lambda_\text{sing} = 2u - 2 \ .
\end{equation}
This implies that the singular vector generates a spin-$j = (u-1)$ representation with basis vector written as $|v^{a_1 \cdots a_{u - 1}}\rangle$ which is totally symmetric and traceless in the indices.

For boundary admissible levels, $u = 2$, $v = 2n + 1$, we have $h = 2n + 1$ and $\lambda = 2$, and $v_\text{sing}$ generates a spin-$1$ representation of $SU(2)$. Therefore for boundary admissible cases, $|\mathcal{N}_T\rangle = K_{ab} J_{-1}^a |v^b\rangle$ of conformal weight $v + 1 = 2n + 2$. The states $|v^a\rangle$ can be written as $J^a_{-1} (K_{ab}J_{-1}^aJ_{-1}^b)^{n}|0\rangle + |c^a_2\rangle$ where $|c_2^a\rangle$ is some $SU(2)$-triplet in the $C_2$ subspace, and $|\mathcal{N}_T\rangle \sim (K_{ab}J_{-1}^aJ_{-1}^b)^{n + 1}|0\rangle + |c_2\rangle$ \cite{Beem:2017ooy}, implementing the nilpotency of stress tensor. For general admissible $\widehat{\mathfrak{su}}(2)_k$ algebras the situation is similar. The states $|v^{a_1 ... a_{u - 1}}\rangle$ can be written as $J_{-1}^{a_1} \ldots J^{a_{u - 1}}_{-1} (K_{ab}J_{-1}^a J_{-1}^b)^{\frac{(u - 1)(v - 1)}{2}}|0\rangle + |c_2^{a_1 ... a_{u - 1}}\rangle$, such that
\begin{equation}
  |\mathcal{N}_T\rangle = L_{-2}^{\frac{(u - 1)(v + 1)}{2}}|0\rangle + |c_2\rangle \ .
\end{equation}
$|\mathcal{N}_T\rangle$ has conformal weight
\begin{equation}
  h[\mathcal{N}_T] = (u - 1)(v + 1)\ .
\end{equation}

Given the above expression of $|\mathcal{N}_T\rangle$ in terms of $|v_\text{sing}\rangle$, we show that it indeed satisfies the $|\mathcal{N}_T\rangle$-conditions (\ref{eq:NT-conditions}). First of all, for boundary admissible $\widehat{\mathfrak{su}}(2)_{- \frac{4n}{2n+1}}$, we have
\begin{equation}
  |v^+\rangle = |v_\text{sing}\rangle, \quad
  |v^1\rangle = - J^-_0|v_\text{sing}\rangle, \quad
  |v^-\rangle = - \frac{1}{2} J_0^-J_0^-|v_\text{sing}\rangle \ .
\end{equation}
It is straightforward to show that $h_{n \ge 2}^1|\mathcal{N}_T\rangle = L_2 |\mathcal{N}_T\rangle = 0$ using $J_{n \ge 1}^a |v_\text{sing}\rangle = 0$. For non-boundary admissible levels and integrable cases, by normal ordering the operators in front of $|v^{b_1 ... b_\ell}\rangle$, one has
\begin{align}
  h_2^1 |\mathcal{N}_T\rangle = & \ - \ell(\ell - 1) K_{a_1 b_1} \cdots K_{a_{\ell - 1}b_{\ell - 1}} J_{-1}^{a_1} \cdots J_{-1}^{a_{\ell - 2}}J_0^{a_{\ell - 1}} |v^{1 b_1 ... b_{\ell - 1}}\rangle \ , \nonumber\\
  h_{n \ge 3}^1 |\mathcal{N}_T\rangle = 0 \\
  L_2K_{a_1 b_1} \cdots K_{a_\ell b_\ell}J_{-1}^{a_1} \cdots & J_{-1}^{a_\ell} |v^{b_1 \cdots b_\ell}\rangle = - \ell K_{a_1 b_1} \cdots K_{a_{\ell}b_{\ell}} J_{-1}^{a_1} \cdots J_{-1}^{a_{\ell - 1}}J_1^{a_{\ell}} |v^{b_1 ... b_{\ell}}\rangle \ . \nonumber
\end{align}
We use the fact that $v^{a_1 ... a_\ell}$ are just several $J_0^-$ acting on $|v_\text{sing}\rangle$, $L_2 |v^{a_1 \cdots }\rangle = 0$, and that
\begin{align}
  h^1_m (J_0^-)^n |v_\text{sing}\rangle
  = & \ - 2n (J_0^-)^{n - 1}J^-_m |v_\text{sing}\rangle + (J_0^-)^n J_m^1 |v_\text{sing}\rangle \ , \\
  J^+_m (J_0^-)^n |v_\text{sing}\rangle
  = & \ - n (n - 1) (J_0^-)^{n - 2} J^-_m |v_\text{sing}\rangle \nonumber \\
  & \  + n (J_0^-)^{n - 1} J_m^1 |v_\text{sing}\rangle + (J_0^-)^n J^+_m |v_\text{sing}\rangle \ ,\\
  J^-_m (J_0^-)^n |v_\text{sing}\rangle = & \ (J_0^-)^n J^-_m |v_\text{sing}\rangle \ .
\end{align}
In particular, $h_m^1 |v^{a_1 \cdots }\rangle = J_m^\pm |v^{a_1 \cdots}\rangle = 0$ for $m \ge 1$ and therefore $L_2 |\mathcal{N}_T\rangle = 0$. Finally we have $h_2^1|\mathcal{N}_T\rangle = 0$ due to
\begin{equation}
  K_{ab} J_0^a |v^{1 b \cdots}\rangle = 0 \ .
\end{equation}

Starting from the $|\mathcal{N}_T\rangle$-equation, one can perform the $S$ transformation. The result is again a sum of lower-weight flavored MLDEs, corresponding to additional null states $(h_1^1)^\ell|\mathcal{N}_T\rangle$.
\begin{equation}
  S|\mathcal{N}_T\rangle = \sum_{\ell \ge 0} \frac{1}{\ell!} \tau^{h[\mathcal{N}_T] - \ell} \mathfrak{b}_{1}^\ell (h^1_1)^\ell |\mathcal{N}_T\rangle \ .
\end{equation}
This formula continues to hold when the starting point is another null such as $|\mathcal{N}\rangle = (h_1^1)^\ell|\mathcal{N}_T\rangle$,
\begin{equation}
  S|\mathcal{N}\rangle = \sum_{\ell \ge 0} \frac{1}{\ell!} \tau^{h[\mathcal{N}] - \ell} \mathfrak{b}_{1}^\ell (h^1_1)^\ell |\mathcal{N}\rangle \ .
\end{equation}
Similar to the discussions on the Deligne-Cvitanovi\'{c} series, it is straightforward to show that $S^2 = 1$. $T$ acts only on $\tau \to \tau + 1$, and can be shown to satisfy $(ST)^3 = 1$. Therefore, $SL(2, \mathbb{Z})$ effectively acts on the null states.

Among all the $SU(2)$ Kac-Moody algebras, only the one with $k = 1$ has a nontrivial singular vector at $h_\text{sing} = 2$, generating a spin-$2$ null representation under $SU(2)$, with basis vector $|v^{ab}\rangle$. There is one charge-zero state $|v^{11}\rangle$ in this representation, and it coincides with the nontrivial state ${(h_1^1)}^2 |\mathcal{N}_T\rangle$ up to some normalization. Hence for $\widehat{\mathfrak{su}}(2)_1$ there are two null states at weight-two that generate two independent flavored modular differential equations: the $|v^{11}\rangle$-equation is in the $S$-orbit of $|\mathcal{N}_T\rangle$-equation, while the Sugawara equation is outside of the $S$-orbit.

For all other $k \ne 1$ cases, the singular vector $|v_\text{sing}\rangle$ generates a suitable null representation as discussed above, which contains a zero charge state $|v^0\rangle$ at conformal weight $h_\text{sing}$. $|v^0\rangle$ is always proportional to a some ${(h_1^1)}^\ell |\mathcal{N}_T\rangle$. Moreover, when $\ell > h[\mathcal{N}_T] - h[v_\text{sing}]$, ${(h_1^1)}^\ell |\mathcal{N}_T\rangle$ vanishes identically, as there is no nontrivial null state except for the descendants of the Sugawara construction $T - T_\text{Sug} = 0$. In particular, ${(h_1^1)}^{h[\mathcal{N}_T] - h[v_\text{sing}]} |\mathcal{N}_T\rangle = 0$ is equivalent to $T - T_\text{Sug} = 0$. As a result, the Sugawara equation for a $k \ne 1$ $SU(2)$ Kac-Moody algebra sits inside the $S$-orbit of $|\mathcal{N}_T\rangle$. The equations from the $S$-orbit of $|\mathcal{N}_T\rangle$ fully determine the spectrum of the algebra.

In the following discussions, we go into some details with simple examples, to provide explicit evidences for the above observations and proposals.

\subsection{\texorpdfstring{$\widehat{\mathfrak{su}}(2)_{2}$}{}}

We take the Kac-Moody algebra $\widehat{\mathfrak{su}}(2)_{2}$ as a non-trivial representative of the integrable algebras. The central charge is $c = 3/2$. The singular vector is given by $v_\text{sing} = (J_{-1}^+)^3|0\rangle$, leading to a spin-$3$ null representation with basis vectors $|v^{abc}\rangle$, with symmetric and traceless indices. The null state $|\mathcal{N}_T\rangle$ is then given by
\begin{equation}
  |\mathcal{N}_T\rangle = \frac{5}{1024} K_{a_1 b_1}K_{a_2 b_2}K_{a_3 b_3} J_{-1}^{b_1}J_{-1}^{b_2}J_{-1}^{b_3} |v^{a_1 a_2 a_3}\rangle \ .
\end{equation}

Through Zhu's recursion, the null state $|\mathcal{N}_T\rangle$ corresponds to the weight-six $|\mathcal{N}_T\rangle$-equation
{\small\begin{align}
  0 = \Bigg[ & \ D_q^{(3)} + \frac{41}{4} E_4(\tau)D_q^{(1)} - \frac{33}{128}E_3[b_1] D_{b_1}^3 \nonumber \\
  & \ + \frac{1}{512} \bigg(
    -1149E_4(\tau)
    + 29 E_2[b_1]^2
    - 586 E_1[b_1]E_3[b_1]
    - 154E_4[b_1]
    \bigg) D_{b_1}^2 \nonumber \\
  & \ + \frac{1}{256} \biggl(
    -9 E_4(\tau)E_1[b_1]
    + 586 E_1[b_1]^2 E_3[b_1]
    - E_1[b_1](29E_2[b_1]^2 + 2162 E_4[b_1])
  \biggr) D_{b_1} \nonumber \\
  & \ + \frac{-4}{256} \biggl(
    198 E_2(\tau) E_3[b_1]
    + 137 E_2[b_1]E_3[b_1]
    + 665E_5[b_1]
  \biggr) -  \nonumber\\
  & \ - \frac{1}{128} \biggl(
    3420 E_6(\tau) + 183 E_4(\tau)E_2[b_1]
    + 29 E_2[b_1]^3
    - 586 E_1[b_1]E_2[b_1]E_3[b_1] \nonumber\\
  & \ \qquad\qquad + 684E_3[b_1]^2
  + 1988 E_2[b_1]E_4[b_1] \nonumber\\
  & \ \qquad \qquad + E_2(\tau)(1149E_4(\tau) - 29E_2[b_1]^2
  + 586E_1[b_1]E_3[b_1]+154E_4[b_1]) \nonumber \\
  & \ \qquad \qquad + 9810E_6[b_1]
  \biggr)\Bigg] \operatorname{ch} \ .
\end{align}}
Here we denote for brevity
\begin{equation}
  E_k[b_1] \coloneqq E_k \begin{bmatrix}
    1 \\ b_1^2
  \end{bmatrix} \ .
\end{equation}

The $|\mathcal{N}_T\rangle$-equation transforms under $S$ into a sum of equations of lower modular weights corresponding to null states of lower conformal weights. In terms of the respective null states,
\begin{equation}
  S|\mathcal{N}_T\rangle = \sum_{\ell \ge 0} \frac{1}{\ell!} \tau^{h[\mathcal{N}_T] - \ell} \mathfrak{b}_{1}^\ell (h^1_1)^\ell |\mathcal{N}_T\rangle \ .
\end{equation}
As discussed above, since the singular vector $|v_\text{sing}\rangle$ is at conformal-weight-three, there is no nontrivial null state at conformal weight-two except for the Sugawara construction. As a result, $(h_1)^4|\mathcal{N}_T\rangle$ must reproduce $T - T_\text{Sug}$. Indeed, $(h_1)^4|\mathcal{N}_T\rangle = 0$. On the other hand, the zero-charge state $|v^{111}\rangle$ in the null representation leads to a nontrivial flavored modular differential equation, which is identified with the state $(h_1)^3|\mathcal{N}_T\rangle$.

The equations in the $S$-orbit of $|\mathcal{N}_T\rangle$-equation determine the entire integrable spectrum of $\widehat{\mathfrak{su}}(2)_2$. With the ansatz
\begin{equation}
  \operatorname{ch} = q^\sigma \sum_{n \ge 0} c_n(b_1)q^n \ ,
\end{equation}
to the leading order the equations impose the following three possible constraints,
\begin{align}
  \sigma = & \ - \frac{1}{16} \ , & c_0' = & \ 0 \ ,\\
  \sigma = & \ + \frac{1}{8} \ , & c_0' = & \ \frac{(b_1^2 - 1)}{b_1(b_1^2 + 1)}c_0 \\
  \sigma = & \ + \frac{7}{16} \ , & \ c_0' = & \ \frac{2(b_1^4 - 1)}{b_1 + b_1^3 + b_1^5} c_0 .
\end{align}
The three solutions correspond to conformal weights $h = 0, \frac{3}{16}, \frac{1}{2}$ respectively, where the first one is the vacuum module. Solving the $c_0$ reveals the finite $\mathfrak{su}(2)$ weights to be
\begin{equation}
  \lambda = 0, \qquad \omega_1, \qquad 2\omega_1 \ .
\end{equation}
These are precisely the highest weights of the integrable modules.

\subsection{\texorpdfstring{$\widehat{\mathfrak{su}}(2)_{- 8/5}$}{}}

The Kac-Moody algebra $\widehat{\mathfrak{su}}(2)_{-8/5}$ is the second simplest boundary admissible algebra with central charge $c = -12$. As a boundary admissible $\widehat{\mathfrak{su}}(2)_k$ algebra, it is the associated chiral algebra of the Argyres-Douglas theory $(A_1, D_5)$. In this case, $k + 2 = \frac{u}{v}$ where $u = 2$, $v = 2n + 1 = 5$, and the singular vector $v_\text{sing}$ is of conformal weight-$5$. Concretely, it reads
\begin{align}
  |v_\text{sing}\rangle = \bigg(& \ -\frac{500}{171} J^3_{-2}J^-_{-1}J^+_{-1}J^+_{-1}+\frac{1250 J^-_{-2}J^+_{-1}J^+_{-1}J^3_{-1}}{1881} \nonumber \\
  & \ -\frac{250}{627} J^+_{-2}J^-_{-1}J^+_{-1}J^3_{-1}-\frac{1250 J^-_{-1}J^+_{-1}J^+_{-1}J^3_{-1}J^3_{-1}}{1881}  \nonumber\\
  & \ -\frac{6070 J^3_{-4}J^+_{-1}}{1881}+\frac{1840 J^3(-3)J^+_{-2}}{1881}+\frac{470}{627} J^+_{-3}J^3_{-2}+\frac{1600}{627} J^3(-3)J^+_{-1}J^3_{-1} \nonumber \\ 
  & \ +\frac{1450 J^+_{-3}J^3_{-1}J^3_{-1}}{1881}-\frac{200}{209} J^+_{-2}J^3_{-2}J^3_{-1}+\frac{500 J^+_{-2}J^3_{-1}J^3_{-1}J^3_{-1}}{1881} \nonumber \\ 
  & \ -\frac{2180 J^+_{-4}J^3_{-1}}{1881}-\frac{7075 J^3_{-2}J^3_{-2}J^+_{-1}}{1881} \nonumber\\
  & \ -\frac{4250 J^3_{-2}J^+_{-1}J^3_{-1}J^3_{-1}}{1881}-\frac{625 J^+_{-1}J^3_{-1}J^3_{-1}J^3_{-1}J^3_{-1}}{1881}-\frac{350}{171} J^-_{-3}J^+_{-1}J^+_{-1} \nonumber \\ 
  & \ +\frac{100}{209} J^-_{-2}J^+_{-2}J^+_{-1}+\frac{2600 J^+_{-3}J^-_{-1}J^+_{-1}}{1881}+\frac{200}{627} J^+_{-2}J^+_{-2}J^-_{-1} \nonumber \\ 
  & \ -\frac{625 J^-_{-1}J^-_{-1}J^+_{-1}J^+_{-1}J^+_{-1}}{1881}+J^+_{-5}\bigg)|0\rangle \ .
\end{align}
Under the action of the finite algebra $\mathfrak{su}(2)$, it generates a spin-$1$ representation $\mathbf{3}$, whose basis may be referred to as $|v^a\rangle$, such that $| v_\text{sing} \rangle = |v^+\rangle$. The $|v^a\rangle$ may be written as
\begin{align}
  |v^a\rangle \propto & \ \big( - \frac{95}{6}J^a_{-1} \mathcal{T}_{-1, -1} \mathcal{T}_{-1,-1} - \frac{8}{5}J_{-5}^a \nonumber \\
  & \ 
  - \frac{5}{4} \mathcal{T}_{-1,-1}\mathcal{T}_{-1,-1}J^a_{-1}
  + \frac{20}{3} \mathcal{T}_{-1,-1}J^a_{-1}\mathcal{T}_{-1,-1} \nonumber \\
  & \ - \frac{77}{3} J^a_{-1}\mathcal{T}_{-3,-1}
  - 9 \mathcal{T}_{-3,-1} J^a_{-1}
  + \frac{11}{3} J^a_{-1} \mathcal{T}_{-2,-2}
  + 8 \mathcal{T}_{-2,-2} J^a_{-1} \nonumber\\
  & \ - 5 \mathcal{T}_{-2,-1} J^a_{-2}
  + \frac{5}{3} J^a_{-2} \mathcal{T}_{-2,-1}\bigg)|0\rangle \ ,
\end{align}
where $\mathcal{T}_{m, n}\coloneqq K_{ab}J_{m}^a J_n^b$.

As proposed above, the null state $|\mathcal{N}_T\rangle$ is of conformal-weight six, related to the singular vector via the simple relation
\begin{equation}
  |\mathcal{N}_T\rangle = K_{ab} J^a_{-1} |v^b\rangle \ .
\end{equation}
More explicitly, we may write
\begin{align}
  |\mathcal{N}_T\rangle = \Big
  ( & \ \frac{2599}{640} K_{a_1a_2}K_{a_3 a_4}K_{a_5a_6}
  - \frac{83}{20}K_{a_1 a_3}K_{a_2 a_5}K_{a_4 a_6}\\
  & \ + \frac{169}{80}K_{a_1 a_4} K_{a_2 a_5} K_{a_3 a_6}
  + \frac{13}{16}K_{a_1 a_3}K_{a_2 a_6}K_{a_4 a_5}\\
  & \ - \frac{11}{64}K_{a_1 a_5} K_{a_2 a_3} K_{a_4 a_6}
  - \frac{503}{160}K_{a_1 a_5} K_{a_2 a_6} K_{a_3 a_4}
  \\
  & \ - \frac{801}{640}K_{a_1 a_6} K_{a_2 a_3} K_{a_4 a_5} + \frac{1179}{320} K_{a_1 a_6} K_{a_2 a_4} K_{a_3 a_5} \Big) J^{a_1}_{-1} ... J^{a_6}_{-1}|0\rangle \ .
\end{align}

Since the first non-trivial singular vector is of conformal-weight $h_\text{sing} = 5$, at lower conformal weights $h = 2, 3, 4$ there are only trivial null states descending from $T - T_\text{Sug}$. As a result,
\begin{align}
  {(h_1^1)}^2|\mathcal{N}_T\rangle = {(h_1^1)}^3|\mathcal{N}_T\rangle = {(h_1^1)}^4|\mathcal{N}_T\rangle = 0 \ .
\end{align}

From the state $|\mathcal{N}_T\rangle$, we extract a flavor MLDE of modular weight-six,
\begin{align}
  0 = & \ \bigg( D_q^3 -10 E_4(\tau) D_q
  + \frac{15}{4} E_3\begin{bmatrix}
    1\\b_1^2
  \end{bmatrix}D_b^3\nonumber\\
  & \ -17 E_4\begin{bmatrix}
    1\\b_1^2
  \end{bmatrix} D_b^2
  +\frac{5}{2} E_2\begin{bmatrix}
    1\\b_1^2
  \end{bmatrix}^2 D_b^2
  + 10 E_1\begin{bmatrix}
    1\\b_1^2
  \end{bmatrix} E_3\begin{bmatrix}
      1\\b_1^2
  \end{bmatrix}D_b^2
  - \frac{75}{4} E_4(\tau) D_b^2\nonumber\\
  & \ -\frac{139}{5} E_5\begin{bmatrix}
      1\\b_1^2
  \end{bmatrix} D_b
  - 83 E_3\begin{bmatrix}
    1\\b_1^2
  \end{bmatrix} E_2\begin{bmatrix}
      1\\b_1^2
  \end{bmatrix}D_b
  - 9 E_4(\tau) E_1\begin{bmatrix}
      1\\b_1^2
  \end{bmatrix} D_b \nonumber\\
  & \ -20 E_1\begin{bmatrix}
    1\\b_1^2
  \end{bmatrix}^2 E_3\begin{bmatrix}
      1\\b_1^2
  \end{bmatrix}D_b \\
  & \ -5 E_1\begin{bmatrix}
    1\\b_1^2
  \end{bmatrix} E_2\begin{bmatrix}
    1\\b_1^2
  \end{bmatrix}^2 D_b
  -59 E_1\begin{bmatrix}
    1\\b_1^2
  \end{bmatrix} E_4\begin{bmatrix}
      1\\b_1^2
  \end{bmatrix}D_b
    -36 E_2 E_3\begin{bmatrix}
      1\\b_1^2
  \end{bmatrix}D_b\nonumber\\
  & \ + 8  E_2\begin{bmatrix}
    1\\b_1^2
  \end{bmatrix}^3
  - 8  E_2 E_2\begin{bmatrix}
    1\\b_1^2
  \end{bmatrix}^2
  +\frac{312}{5}  E_4 E_2\begin{bmatrix}
    1\\b_1^2
  \end{bmatrix}+32  E_1\begin{bmatrix}
      1\\b_1^2
  \end{bmatrix} E_3\begin{bmatrix}
      1\\b_1^2
  \end{bmatrix} E_2\begin{bmatrix}
      1\\b_1^2
  \end{bmatrix} \nonumber\\
  & \ +\frac{232}{5}  E_4\begin{bmatrix}
      1\\b_1^2
  \end{bmatrix} E_2\begin{bmatrix}
      1\\b_1^2
  \end{bmatrix}+\frac{864}{5}  E_3\begin{bmatrix}
    1\\b_1^2
  \end{bmatrix}^2 -32  E_2 E_1\begin{bmatrix}
      1\\b_1^2
  \end{bmatrix} E_3\begin{bmatrix}
      1\\b_1^2
  \end{bmatrix}\nonumber\\
  & \ +\frac{272}{5}  E_2 E_4\begin{bmatrix}
      1\\b_1^2
  \end{bmatrix}
  -\frac{504}{5}  E_6\begin{bmatrix}
      1\\b_1^2
  \end{bmatrix}
  + 60  E_2 E_4-\frac{252  E_6}{5} \bigg) \operatorname{ch} \ . \nonumber
\end{align}

Under the modular $S$-transformation,
\begin{equation}
  \tau \to - \frac{1}{\tau}, \qquad \mathfrak{b}_1 \to \frac{\mathfrak{b}_1}{\tau}, \qquad
  \mathfrak{y} \to \mathfrak{y} - \frac{1}{\tau} \mathfrak{b}_1^2 \ ,
\end{equation}
the above weight-six equation transforms into the following structure,
\begin{equation}
  S |\mathcal{N}_T\rangle = \sum_{\ell=0}^{4} \frac{1}{\ell!}\tau^{h[\mathcal{N}_T]-\ell} \mathfrak{b}_1^\ell (h^1_{1})^\ell |\mathcal{N}_T\rangle \ .
\end{equation}
In particular, the $\tau^2 \mathfrak{b}_1^4$ term reads (up to a normalization factor $\frac{192}{25}$)
\begin{equation}
  0 = D_q^{(1)}\operatorname{ch} - \frac{5}{8}D_{b_1}^2 \operatorname{ch}
  - \frac{5}{2} E_1 \begin{bmatrix}
    1\\b_1^2
  \end{bmatrix} D_{b_1}\operatorname{ch}
  + 4E_2 \begin{bmatrix}
    1 \\ b_1^2
  \end{bmatrix}\operatorname{ch}
  + 2E_2(\tau) \operatorname{ch} \ ,
\end{equation}
which is simply the equation following from the Sugawara construction $T = T_\text{Sug} $. Similarly, take any of the null states $|\mathcal{N}\rangle = {(h_1^1)}^\ell|\mathcal{N}_T\rangle$ and the corresponding equation, the $S$-transformation continue to yield the same structure,
\begin{equation}
  S|\mathcal{N}\rangle = \sum_{\ell\ge 0} \frac{1}{\ell!}\tau^{h[\mathcal{N}]-\ell} \mathfrak{b}_1^\ell (h^1_{1})^\ell |\mathcal{N}\rangle \ .
\end{equation}

Now we may collect the $S$-orbit of $|\mathcal{N}_T\rangle$-equation and find their common solutions. Under the ansatz
\begin{equation}
  \operatorname{ch} = q^\sigma \sum_{n \ge 0} c_n(b_1) q^n \ , 
\end{equation}
to the leading order in $q$, the equations require
\begin{align}
  \begin{array}{ll}
    \displaystyle\sigma = - \frac{1}{10}, & \displaystyle c''_0 = - \frac{24}{25b_1^2}c_0 - \frac{1 + 3 b_1^2}{b_1(b_1^2 - 1)}c_0' \ , \\
    \displaystyle\sigma = + \frac{1}{10}, & \displaystyle c_0''(b_1) = - \frac{16}{25b_1^2}c_0 - \frac{(1 + 3 b_1^2)}{b_1(b_1^2 - 1)}c'_0 \ , \\
    \displaystyle\sigma = + \frac{1}{2},  & c_0' = 0 \ .
  \end{array}
\end{align}
The last solution clearly corresponds to the vacuum module with with $- \frac{1}{24}c = \frac{1}{2}$. The remaining two solutions can be solved easily, giving the conformal weight $h$ and $SU(2)$ weight of the module,
\begin{align}
  h = & \ - \frac{2}{5}  & c_0 = & \ C_1 \frac{b_1^{-6/5}}{1 - b_1^{-2}} + C_2 \frac{b_1^{-4/5}}{1 - b_1^{-2}} \\
  h = & \ - \frac{3}{5} & c_0 = & \ C_1 \frac{b_1^{- 8/5 }}{1 - b_1^{-2}} + C_2 \frac{b_1^{- 2/5}}{1 - b_1^{-2}} \ .
\end{align}
The first line recovers the admissible modules with finite $\mathfrak{su}(2)$ highest weight $\lambda = - \frac{6}{5} \omega_1$ or $\lambda = - \frac{4}{5}\omega_1$, while the second line recovers $\lambda = - \frac{8}{5}\omega_1$ or $\lambda = - \frac{2}{5}\omega_1$. Together with the vacuum module, we precisely recover all the admissible modules of $\widehat{\mathfrak{su}}(2)_{-8/5}$.

\subsection{\texorpdfstring{$\widehat{\mathfrak{su}}(2)_{-1/2}$}{}}

The $\widehat{\mathfrak{su}}(2)$ algebra with level $k = - \frac{1}{2}$ is admissible but not boundary admissible, with central charge $c = -1$. The singular vector is of weight-four,
\begin{align}
  |v_\text{sing}\rangle \propto \Big(& \frac{13}{2} J^3_{-2}J^+_{-1}J^+_{-1}-\frac{3}{2} J^+_{-2}J^+_{-1}J^3_{-1}+J^+_{-1}J^+_{-1}J^3_{-1}J^3_{-1} \nonumber \\ 
  & \ \qquad +J^-_{-1}J^+_{-1}J^+_{-1}J^+_{-1}-\frac{31}{4} J^+_{-3}J^+_{-1}+\frac{15}{16} J^+_{-2}J^+_{-2} \Big)|0\rangle \ .
\end{align}
It generate a spin-$2$ representation under the finite $\mathfrak{su}(2)$, whose basis can be written as
\begin{align}
  |v^{ab}\rangle = \left({\frac{17}{16}J^a_{-1}J^b_{-1}\mathcal{T}_{-1,-1}
  - \frac{3}{4}J^a_{-1}\mathcal{T}_{-1,-1}J^b_{-1}
  + \frac{3}{16} \mathcal{T}_{-1,-1}J^a_{-1}J^b_{-1} + \frac{3}{16} J^a_{-2 }J^b_{-2}}\right)|0\rangle \nonumber  \\
  + (a \leftrightarrow b) - \operatorname{trace} \nonumber \ ,
\end{align}
where $\mathcal{T}_{-1,-1}\coloneqq K_{ab}J^a_{-1}J^b_{-1}$, the removal of trace is to ensure $K_{ab}v^{ab} = 0$. The null state $|\mathcal{N}_T\rangle$ is of conformal weight-six and is related to the singular vector simply by
\begin{equation}
  |\mathcal{N}_T\rangle = \frac{1}{9} K_{a a'}K_{b b'}J_{-1}^aJ_{-1}^b |v^{a'b'}\rangle \ .
\end{equation}

The flavored MLDE corresponding to $|\mathcal{N}_T\rangle$ reads
\begin{align}\label{eq:NT-su2-1/2}
0 = & \ \Bigg[\frac{4}{5} \Bigg(10 E_2 \begin{bmatrix}
    1\\b_1^2
\end{bmatrix}^3+78 E_4 E_2 \begin{bmatrix}
    1\\b_1^2
\end{bmatrix}+40 E_1 \begin{bmatrix}
    1\\b_1^2
\end{bmatrix} E_3 \begin{bmatrix}
    1\\b_1^2
\end{bmatrix} E_2 \begin{bmatrix}
    1\\b_1^2
\end{bmatrix} \nonumber\\ 
& \ +58 E_4 \begin{bmatrix}
    1\\b_1^2
\end{bmatrix} E_2 \begin{bmatrix}
    1\\b_1^2
\end{bmatrix}+216 \left(E_3 \begin{bmatrix}
    1\\b_1^2
\end{bmatrix}\right)^2 \nonumber\\
& \ +E_2 \bigg(-10 E_2 \begin{bmatrix}
  1\\b_1^2
\end{bmatrix}^2
-40 E_1 \begin{bmatrix}
    1\\b_1^2
\end{bmatrix} E_3 \begin{bmatrix}
    1\\b_1^2
\end{bmatrix}+68 E_4 \begin{bmatrix}
    1\\b_1^2
\end{bmatrix}+75 E_4\bigg)\nonumber \\
& \ \qquad -126 E_6 \begin{bmatrix}
    1\\b_1^2
\end{bmatrix}
-63 E_6\Bigg)\nonumber \\
& \ +\Bigg(-20 E_3 \begin{bmatrix}
    1\\b_1^2
\end{bmatrix} E_1 \begin{bmatrix}
  1\\b_1^2
\end{bmatrix}^2-9 E_4 E_1 \begin{bmatrix}
    1\\b_1^2
\end{bmatrix}- \Big(5 E_2 \begin{bmatrix}
  1\\b_1^2
\end{bmatrix}^2+59 E_4 \begin{bmatrix}
    1\\b_1^2
\end{bmatrix} \Big) E_1 \begin{bmatrix}
    1\\b_1^2
\end{bmatrix} \nonumber \\
& \ \qquad\quad -36 E_2 E_3 \begin{bmatrix}
    1\\b_1^2
\end{bmatrix}-83 E_2 \begin{bmatrix}
    1\\b_1^2
\end{bmatrix} E_3 \begin{bmatrix}
    1\\b_1^2
\end{bmatrix}-\frac{139}{5} \text{EE}(5) \begin{bmatrix}
    1\\b_1^2
\end{bmatrix}\Bigg) D_b \nonumber \\
& \ +\left(\frac{5}{2} \left(E_2 \begin{bmatrix}
    1\\b_1^2
\end{bmatrix}\right)^2+10 E_1 \begin{bmatrix}
    1\\b_1^2
\end{bmatrix} E_3 \begin{bmatrix}
    1\\b_1^2
\end{bmatrix}-17 E_4 \begin{bmatrix}
    1\\b_1^2
\end{bmatrix}-\frac{75 E_4}{4}\right)D_b^2 \nonumber \\
& \ +\frac{15}{4} E_3 \begin{bmatrix}
    1\\b_1^2
\end{bmatrix}D_b^3 -10 E_4 D_q^{(1)} + D_q^{(3)} \Bigg]\operatorname{ch} \ .
\end{align}

Under the $S$-transformation
\begin{equation}
  \tau \to - \frac{1}{\tau}, \qquad \mathfrak{b}_1 \to \frac{\mathfrak{b}_1}{\tau}, \qquad \mathfrak{y} \to \mathfrak{y} - \frac{1}{\tau} \mathfrak{b}_1^2 \ ,
\end{equation}
the above weight-six equation transforms into a sum of equations of lower weights. In terms of the corresponding null states, we have
\begin{equation}
  S |\mathcal{N}_T\rangle = \sum_{\ell=0}^{4} \frac{1}{\ell!}\tau^{6-\ell} \mathfrak{b}_1^\ell (h^1_{1})^\ell |\mathcal{N}_T\rangle \ .
\end{equation}
Explicitly, we list here the equations that appear in the transformation. The weight-five null state $h^1_1|\mathcal{N}_T\rangle$ corresponds to the equation
\begin{align}\label{eq:weight-5-su2-1/2}
  0 = \Bigg[ & \ 3 D_{b_1} D_q^{(2)} + 3 E_2(\tau) D_{b_1} D_q^{(1)}
  + \frac{3}{4} E_2 \begin{bmatrix}
    1 \\ b_1^2
  \end{bmatrix} D_{b_i}^3
  \nonumber\\
  & \ +  \bigg(3 E_1 \begin{bmatrix}
    1 \\ b_1^2
  \end{bmatrix}E_2 \begin{bmatrix}
    1 \\ b_1^2
  \end{bmatrix} + \frac{39}{4} E_3 \begin{bmatrix}
    1 \\b_1^2
  \end{bmatrix} \bigg)D_{b_1}^2\nonumber\\
  & \ + \bigg(8 E_2(\tau)^2 - \frac{155}{8}E_4(\tau)
  + 2 q\partial_q E_2(\tau) - \frac{9}{4}E_2(\tau)E_2 \begin{bmatrix}
    1 \\ b_1^2
  \end{bmatrix} - 6 E_1 \begin{bmatrix}
    1 \\ b_1^2
  \end{bmatrix}^2 E_2 \begin{bmatrix}
    1 \\ b_1^2
  \end{bmatrix}\nonumber\\
  & \ \qquad -6 E_2 \begin{bmatrix}
    1 \\ b_1^2
  \end{bmatrix}^2 + 30 E_1 \begin{bmatrix}
    1 \\ b_1^2
  \end{bmatrix}E_3 \begin{bmatrix}
    1 \\ b_1^2
  \end{bmatrix} - \frac{225}{8}E_4 \begin{bmatrix}
    1 \\ b_1^2
  \end{bmatrix} \bigg) D_{b_1}\nonumber\\
  & \ + \frac{1}{12} \biggl(
    102 E_4(\tau) E_1 \begin{bmatrix}
      1 \\ b_1^2
    \end{bmatrix}
    + 36 E_1 \begin{bmatrix}
      1 \\ b_1^2
    \end{bmatrix}
    E_2 \begin{bmatrix}
      1 \\ b_1^2
    \end{bmatrix}^2
    + 232 E_2 \begin{bmatrix}
      1 \\ b_1^2
    \end{bmatrix}E_3 \begin{bmatrix}
      1 \\ b_1^2
    \end{bmatrix} \nonumber\\
  & \ \qquad \qquad - 9 E_2(\tau) \Big(4E_1 \begin{bmatrix}
    1 \\ b_1^2
  \end{bmatrix}E_2 \begin{bmatrix}
    1 \\ b_1^2
  \end{bmatrix}
  + 13 E_3 \begin{bmatrix}
    1 \\ b_1^2
  \end{bmatrix} \Big) \nonumber\\
  & \ \qquad\qquad - 102 E_1 \begin{bmatrix}
    1 \\ b_1^2
  \end{bmatrix}E_4 \begin{bmatrix}
    1 \\ b_1^2
  \end{bmatrix}
  + 775 E_5 \begin{bmatrix}
    1 \\ b_1^2
  \end{bmatrix}
  \biggr)\Bigg] \operatorname{ch} \ .
\end{align}
The weight-four state $\frac{1}{2}h_1^1h_1^1|\mathcal{N}_T\rangle$ corresponds to the weight-four flavored equation
\begin{align}\label{eq:weight-4-su2-1/2}
  \Bigg[& - \frac{3}{2}D_q^{(2)} + 3 D_{b_1}^2 D_q^{(1)} - 3 E_2(\tau) D_q^{(1)}
  - \frac{3}{4} E_1 \begin{bmatrix}
    1 \\ b_1^2
  \end{bmatrix} D_{b_1}^3 \nonumber\\
  & \ + \bigg(6E_2(\tau) - 3 E_1 \begin{bmatrix}
    1 \\ b_1^2
  \end{bmatrix}^2
  - \frac{63}{8}E_2 \begin{bmatrix}
    1 \\ b_1^2
  \end{bmatrix} \bigg) D_{b_1}^2 \nonumber \\
  & \ + \bigg(\frac{9}{4}E_2(\tau) E_1 \begin{bmatrix}
    1 \\ b_1^2
  \end{bmatrix}
  + 6 E_1 \begin{bmatrix}
    1 \\ b_1^2
  \end{bmatrix}^3
  + 27 E_1 \begin{bmatrix}
    1 \\ b_1^2
  \end{bmatrix}E_2 \begin{bmatrix}
    1 \\ b_1^2
  \end{bmatrix}
  + \frac{387}{8}E_3 \begin{bmatrix}
    1 \\ b_1^2
  \end{bmatrix}
  \bigg) D_{b_1} \nonumber\\
  & \ - \frac{1}{48}\biggl(
    144E_2(\tau)^2
    - 633 E_4(\tau)
    + 144 E_1 \begin{bmatrix}
      1 \\ b_1^2
    \end{bmatrix}^2E_2 \begin{bmatrix}
      1 \\ b_1^2
    \end{bmatrix}
    - 80 E_2 \begin{bmatrix}
      1 \\ b_1^2
    \end{bmatrix}
  \nonumber\\
  & \ \qquad \qquad - 18 E_2 (\tau) \biggl(
    8 E_1 \begin{bmatrix}
      1 \\ b_1^2
    \end{bmatrix}^2
    + 21 E_2 \begin{bmatrix}
      1 \\ b_1^2
    \end{bmatrix}\biggr) \nonumber\\
    & \ \qquad \qquad + 1240 E_1 \begin{bmatrix}
      1 \\ b_1^2
    \end{bmatrix}E_3 \begin{bmatrix}
      1 \\ b_1^2
    \end{bmatrix}
    + 2833 E_4 \begin{bmatrix}
      1 \\ b_1^2
    \end{bmatrix}\biggr) 
  \Bigg]\operatorname{ch} \ .
\end{align}
The weight-three state $\frac{1}{3!}(h_1^1)^3|\mathcal{N}_T\rangle$ corresponds to the equation
\begin{align}\label{eq:weight-3-su2-1/2}
  \Bigg[&  -3 D_{b_1}D_q^{(1)} + \frac{1}{2}D_{b_1}^3 + 2 E_1 \begin{bmatrix}
    1 \\ b_1^2
  \end{bmatrix} D_{b_1}^2 
  + \bigg(- \frac{9}{2}E_2(\tau) - 4 E_1 \begin{bmatrix}
    1 \\ b_1^2
  \end{bmatrix}^2 - 9 E_2 \begin{bmatrix}
    1 \\ b_1^2
  \end{bmatrix}\bigg) D_{b_1} \nonumber\\
  & \ - 2 E_2 (\tau)E_1 \begin{bmatrix}
    1 \\ b_1^2
  \end{bmatrix}
  +2  E_1 \begin{bmatrix}
    1 \\ b_1^2
  \end{bmatrix}E_2 \begin{bmatrix}
    1 \\ b_1^2
  \end{bmatrix}
  + 6 E_3 \begin{bmatrix}
    1 \\ b_1^2
  \end{bmatrix}
  \Bigg] \operatorname{ch} = 0 \ .
\end{align}
Finally, the weight-two state $\frac{1}{4!}(h_1^1)^4|\mathcal{N}_T\rangle$ corresponds to the equation
\begin{equation}\label{eq:weight-2-su2-1/2}
  \Bigg[\frac{3}{4}D_q^{(1)} - \frac{1}{8} D_{b_1}^2 + \frac{1}{2}E_1 \begin{bmatrix}
    1 \\ b_1^2
  \end{bmatrix}D_{b_1} + \frac{1}{8}\bigg(E_2(\tau) + 2 E_2 \begin{bmatrix}
    1 \\ b_1^2
  \end{bmatrix} \bigg) \Bigg]\operatorname{ch} = 0 \ .
\end{equation}
We recognize this equation to be simply the Sugawara equation, up to a factor of $3/4$. In other words, the Sugawara condition $T - T_\text{Sug} = 0$ sits within the modular orbit of the $|\mathcal{N}_T\rangle$. We also remark that both ${(h_1^1)}^3|\mathcal{N}_T\rangle$ and ${(h_1^1)}^4|\mathcal{N}_T\rangle$ are both zero trivially, but through Sugawara conditions, they do give rise to non-trivial flavored MLDEs. On the other hand, for $(h^1_1)^5|\mathcal{N}_T\rangle = 0$ and $(h^1_1)^6 |\mathcal{N}_T\rangle = 0$ there are no corresponding nontrivial flavored MLDEs.

One can work out the modular $S$-transformation of any of the above equations, which can be summarized in terms of the corresponding null states,
\begin{equation}
  S |\mathcal{N}\rangle = \sum_{\ell \ge 0} \frac{1}{\ell!}\tau^{h[\mathcal{N} ]- \ell} \mathfrak{b}_1^\ell {(h_1^1)}^\ell |\mathcal{N}\rangle \ ,
\end{equation}
where $|\mathcal{N}\rangle$ denotes the null state corresponding to the equation in question.

Now we are ready to recover the spectrum of the algebra by computing all the characters as a $q$-series. We write
\begin{equation}
  \operatorname{ch}(q, b_1) = q^{\sigma} \sum_{n = 0}^{+\infty}a_n(b_1)q^n \ .
\end{equation}
Applying the above equations from weight-2 to weight-6 to the leading order to the ansatz, we may determine the possible values of $\sigma$ and the differential equation that governs the leading functional coefficient $a_0(b_i)$,
\begin{equation}
  \left\{\begin{array}{ll}
    \displaystyle\sigma = + \frac{1}{24} & a'(b_1) = 0 \\
    \displaystyle \sigma = + \frac{13}{24} & \displaystyle a'_0(b_1) = \frac{(b_1^2 - 1)}{b_1 (1 + b_1^2)}a_0(b_1) \\
    \displaystyle\sigma = - \frac{1}{12}, & \displaystyle a'_0(b_1) = \frac{(b_1^2 - 1)(3 a_0(b_1) + 4 b_1^2 c''_0(b_1))}{4(b_1 + 3 b_1^3)}
  \end{array}\right.\ .
\end{equation}
The first solution corresponds to the vacuum representation with conformal weight $h = 0$. The second solution gives
\begin{equation}
  a_0(b_1) = b_1 + b_1^{-1} \ ,
\end{equation}
and it corresponds to an ordinary modules $h = 1/2$, with the finite highest weight $\omega_1$. The third solution gives two solutions,
\begin{equation}
  a_0(b_1) = \frac{c_1b_1^{-3/2} + c_2 b_1^{-1/2}}{1 - b_1^{-2}} \ .
\end{equation}
These two solutions correspond to non-ordinary modules with finite highest weights $- \frac{1}{2}\omega_1$ and $- \frac{3}{2}\omega_1$ as indicated by the powers of $b_1$ in the numerator. The remaining coefficients can be algebraically solved using the weight-6 and weight-2 equations. For example,
\begin{align}
  a_1(b_1) = 
  & \ - \frac{144 (-27(b_1^4 + 1) + (576\alpha^2 + 168\alpha - 35)b_1^2)}{(13 + 12\alpha)(11 + 24 \alpha)(23 + 24\alpha)}a_0(b_1) \nonumber \\
  & \ + \frac{432 \left(30 b_1^6+125 b_1^4+114 b_1^2+11\right) }{(12 \alpha +13) (24 \alpha +11) (24 \alpha +23) (b_1-1) b_1 (b_1+1)} a_0'(b_1) \nonumber\\ 
  & \ + \frac{216 \left(b_1^4+30 b_1^2+37\right) }{(12 \alpha +13) (24 \alpha +11) (24 \alpha +23)}a_0''(b_1) \nonumber\\
  & \ - \frac{1296 (b_1-1) b_1 (b_1+1) \left(b_1^2+1\right) }{(12 \alpha +13) (24 \alpha +11) (24 \alpha +23)} a_0'''(b_1)\ .
\end{align}
All higher order terms $a_n(b_1)$ can be solved in a similar manner once $a_0, ..., a_{n - 1}$ are known. These solutions precisely recover all the admissible module characters of the $\widehat{\mathfrak{su}}(2)_{-1/2}$ algebra.

\subsection{\texorpdfstring{$\mathcal{N} = 4$ small superconformal algebra}{}}

As a preliminary extension of the previous discussions, we consider the 2d small $\mathcal{N} = 4$ superconformal algebra. It is the associated chiral algebra of the 4d $\mathcal{N} = 4$ super Yang-Mills theory with an $SU(2)$ gauge group. The algebra has bosonic generators $J^a$, $T(z)$ and fermionic generators $G_\pm$, $\tilde G_\pm$. The currents $J^a$ form an $\widehat{\mathfrak{su}}(2)_{-3/2}$ subalgebra, and the Sugawara stress tensor $T_\text{Sug}$ happens to be the stress tensor $T$ of the full algebra. The commutation relations between $T, J$ are as usual, while those involving supercharges $G, \tilde G$ are given by \cite{1976PhLB...62..105A,Beem:2017ooy,Kos:2018glc}
\begin{align}
  \{G^\alpha_m, \tilde G^\beta_n\} = & \ - \epsilon^{\alpha\beta}L_{m + n} + (m - n) K_{ab}(\sigma^a)^{\alpha \beta} J_{m + n}^b - \frac{c}{6}(m^2 - \frac{1}{4}) \epsilon^{\alpha \beta} \delta_{m + n, 0} \ , \nonumber \\
  [J_m^a, G^\alpha_{n}] = & \ (\sigma^a)_\beta {^\alpha}G^\beta_{m + n}, \qquad
  [J_m^a, \tilde G^\alpha_{n}] =  (\sigma^a)_\beta {^\alpha}\tilde G^\beta_{m + n} \ ,\\ 
  [L_m, G^\alpha_{n}] = & \ (m h_G - m - n)G^\alpha_{m + n}, \qquad
  [L_m, \tilde G^\alpha_n] = (m h_G - m - n)\tilde G^\alpha_{m + n} \ . \nonumber
\end{align}
Here $(\sigma^a)_\alpha{^\beta}$ denotes the standard Pauli matrices, $(\sigma^a)^{\alpha \beta} = \epsilon^{\alpha \alpha'}(\sigma^a)_{\alpha'}{^\beta}$, $\sigma^\pm = \frac{1}{2}(\sigma^1 \pm i \sigma^2)$, $\epsilon^{12} = - \epsilon^{21} = - \epsilon_{12} = - 1$, $K_{ab}$ is the $\mathfrak{su}(2)$ Killing form.

As shown in \cite{Beem:2017ooy}, there are several null states worth noting, some of which are related to chiral ring relations in four dimensions. We focus on three null states that correspond to flavored modular differential equations,
\begin{align}
  |\mathcal{N}_\text{Sug}\rangle = & \ \Big(L_{-2} - \frac{1}{2(k + h^\vee)}K_{ab}J^a_{-1}J^b_{-1} \Big)|0\rangle \ , \nonumber \\
  |\mathcal{N}_{G\tilde G}^1\rangle = & \ (2L_{-2}J^a_{-1} - \sigma^a_{\alpha \beta}G^\alpha_{-3/2}\tilde G^\beta_{-3/2}
  - K^{bb'} f^{ab}{_c}J_{-2}^{b'}J_{-1}^{c} - 4J^a{_{-3}})|0\rangle \ , \\ 
  |\mathcal{N}_T\rangle = & \ \Big(L_{-2}^2 + \frac{1}{2} \epsilon_{\alpha \beta} (\tilde G^\alpha_{-5/2}G^\beta_{-3/2} + G^\alpha_{-5/2}\tilde G^\beta_{-3/2})
  - \frac{1}{2} K_{ab}J_{-2}^aJ_{-2}^b - L_{-4} \Big)|0\rangle \ . \nonumber
\end{align}
The last null state leads to weight-four $\mathcal{N}_T$-equation
\begin{align}
  0 = \Bigg[ D_q^{(2)} & \ - 2E_2\begin{bmatrix}
    -1 \\ b_1
  \end{bmatrix}D_q^{(1)} 
  + \biggl(2 E_3 \begin{bmatrix}
    1 \\ b_1^2
  \end{bmatrix} - 4 E_3 \begin{bmatrix}
    -1 \\ b_1
  \end{bmatrix}\biggr)D_{b_1} \nonumber \\
  & \ + \bigg(6k E_4 \begin{bmatrix}
    1 \\ b_1^2
  \end{bmatrix} - 2cE_4 \begin{bmatrix}
    -1 \\ b_1
  \end{bmatrix} + \frac{c}{2}E_4(\tau) + 3k E_4(\tau) \bigg)\Bigg]\operatorname{ch} \nonumber
\end{align}
The modular group of this equation is $\Gamma^0(2)$ instead of $SL(2, \mathbb{Z})$. Under the modular $STS \in \Gamma^0(2)$ transformation, it transforms into a sum of equations associated to $\mathcal{N}_T$, $\mathcal{N}_{G \tilde G}^1, \mathcal{N}_{G\tilde G}$ \cite{Zheng:2022zkm}. Note that by direct computation,
\begin{equation}
  |\mathcal{N}_{G \tilde G}^1\rangle = h_1^1|\mathcal{N}_T\rangle, \qquad
  (h_1^1)^2 |\mathcal{N}_T\rangle = - 4 \biggl(L_{-2} - \frac{1}{2(k + h^\vee)}K_{ab}J^a_{-1}J^b_{-1}\biggr)|0\rangle
\end{equation}
Moreover, one can check that $\mathcal{N}_{G \tilde G}^1$ satisfies $J^a_{n \ge 1}|\mathcal{N}_{G\tilde G}^1\rangle = J^+_0|\mathcal{N}_{G\tilde G}^1\rangle = 0$, and is a singular vector with respect to $\widehat{\mathfrak{su}}(2)_{-3/2}$ subalgebra. Under the finite $\mathfrak{su}(2)$ algebra, $|\mathcal{N}_{G \tilde G}^+\rangle$ generates a spin-$1$ representation with basis $|\mathcal{N}_{G\tilde G}^a\rangle = J^a_1 |\mathcal{N}_T\rangle$. $|\mathcal{N}_T\rangle$ is a descendant of $|\mathcal{N}_{G\tilde G}^+\rangle$, and satisfies the $|\mathcal{N}_T\rangle$-constraint,
\begin{equation}
  L_2 |\mathcal{N}_T\rangle = J^a_{n \ge 2}|\mathcal{N}_T\rangle = 0 \ .
\end{equation}
In fact, one may consider the most general state $|\mathcal{N}_T\rangle$ at weight $h = 4$ that implement the nilpotency of $T$,
\begin{align}
  |\mathcal{N}_T\rangle
  = \Big( &  L_{-2}^2 + \delta L_{-4}  + \beta K_{ab}J^a_{-3}J^b_{-1} + \gamma K_{ab}J^a_{-2}J^b_{-2}\\
  & \ + \delta_1 \epsilon_{\alpha \beta}G^\alpha_{-5/2}\tilde G^\beta_{-3/2}
  + \delta_2 \epsilon_{\alpha \beta}\tilde G^\alpha_{-5/2} G^\beta_{-3/2}\Big)|0\rangle \ .
\end{align}
Applying the $|\mathcal{N}_T\rangle$-constraint together with quasi-modularity for Kac-Moody algebra on selects three nontrivial $k$: $k = -3/2, -4/3, 1$ \footnote{There is an additional solution with $c = -16$ to the $|\mathcal{N}_T\rangle$-conditions. However, this solution fails the quasi-modularity test: the $\tau^0$ and $\tau^1$ terms in the $S$-transformed $|\mathcal{N}_T\rangle$-equation do not give consistent solution for $k$.}. We recognize the latter two solutions to be two simplest cases in section 2, while the first one is precisely the level in the genuine small $\mathcal{N} = 4$ superconformal algebra with $c = -9$. This simply means that the $|\mathcal{N}_T\rangle$-conditions should be further enhanced to accommodate the additional generators $G, \tilde G$ of the algebra. One possible enhancement may be
\begin{equation}
  G^\alpha_{5/2}|\mathcal{N}_T\rangle = \tilde G^\alpha_{5/2}|\mathcal{N}_T\rangle = 0 \ .
\end{equation}
This additional condition singles out $k = -3/2$. We leave a more systematic quasi-modular bootstrap outside of realm of Kac-Moody algebras for future work.

\section{Translation and flavored modular differential equations}

\subsection{Untwisted spectral flow and null states}

For a Kac-Moody algebra $\widehat{\mathfrak{g}}_k$, one can define the spectral flow automorphisms $\tau_i$ that rearrange generators $J^a_n$ and the Virasoro operator $L_0$ depending on the simple coroot $\alpha^\vee_i$ \cite{Ridout:2008nh}. In particular, starting from the vacuum module of an admissible Kac-Moody algebra, one can perform suitable spectral flow to obtain another admissible module.

As discussed above, the module characters the Kac-Moody algebras in the Deligne-Cvitanovi\'c series and all admissible $\widehat{\mathfrak{su}}(2)_k$ algebras solve a set of flavored modular differential equations. Therefore, it is natural to expect that the spectral flow automorphisms are somehow encoded in the equations. Indeed, for the Deligne-Cvitanovi\'c series, under the translation $\sigma$ depending on rationals $n_i \in \mathbb{Q}$,
\begin{align}
  & \ \mathfrak{b}_i \xrightarrow{\sigma} \mathfrak{b}_i + n_i \tau, \qquad
  \mathfrak{y} \xrightarrow{\sigma} \mathfrak{y} + \sum_{i,j=1}^{r}K^{ij}n_j \mathfrak{b}_i
  + \sum_{i,j = 1}^{r} \frac{1}{2}K^{ij} n_i n_j \tau \ , \\
  & \ (\mathbf{n}^\vee, \alpha) \in \mathbb{Z}, \qquad \mathbf{n}^\vee \coloneqq \sum_{i = 1}^r n_i \alpha^\vee_i \ ,\qquad \forall \alpha \in \Delta,
\end{align}
the corresponding flavored modular differential equations transform among themselves, in a way similar to the quasi-modularity \cite{Pan:2023jjw}. These translation can be used to obtain non-vacuum module characters from the vacuum character \footnote{This does not mean that the translation equals spectral flow operation, since a translation does not always give the correct map between states. However our discussions concern mainly the characters as analytic functions, hence shall ignore this caveat.}.

We now generalize this behavior under translation to all integrable and admissible $\widehat{\mathfrak{su}}(2)_k$ Kac-Moody algebras. In particular, we focus on the $|\mathcal{N}_T\rangle$ equation. In the $\mathfrak{su}(2)$ case, $r = 1$, $K^{11} = 2$, we find that under the above translation (where $n_1$ may take half-integral values),
\begin{equation}
  \mathfrak{b}_1 \xrightarrow{\sigma} \mathfrak{b}_1 + n_1 \tau, \qquad \mathfrak{y} \xrightarrow{\sigma} \mathfrak{y} + 2n_1 \mathfrak{b}_1 + n_1^2 \tau\ ,
\end{equation}
the $|\mathcal{N}_T\rangle$-equation transforms into those equations of equal or lower weight. Note that the translation is identical to the spectral flow discussed in \cite{2023arXiv230409681L} if we identify $n_1 = \frac{\ell}{2}$ there. Also note that $b_1$ enters into the equation by $b_1^2$, which transforms to $b_1^2 q^{2n_1} = b_1^2 q^\ell$ under translation, hence as long as $\ell \in \mathbb{Z}$, the twist parameter $+1$ in the Eisenstein series remain unchanged. In terms of the corresponding null states, the translation on the $|\mathcal{N}_T\rangle$-equation can be elegantly written as
\begin{align}\label{eq:translation-su2-untwisted}
  |\mathcal{N}_T\rangle \xrightarrow{\sigma} \sum_{\ell \ge 0} \frac{(-1)^\ell}{\ell!}n_1^\ell (h_1^1)^\ell |\mathcal{N}_T\rangle \ .
\end{align}

For example, consider $\widehat{\mathfrak{su}}(2)_{-1/2}$. It has four admissible modules, with (the finite part of) the highest weight $\lambda = 0, + \omega_1, - \frac{3}{2}\omega_1, - \frac{1}{2}\omega_1$. The flavored characters are given by \cite{Ridout:2008nh}
\begin{align}
  \operatorname{ch}_0 = & \ \frac{1}{2} \bigg[
    \frac{\eta(\tau)}{\vartheta_4(\mathfrak{b}_1)}
    + \frac{\eta(\tau)}{\vartheta_3(\mathfrak{b}_1)}
    \bigg], \qquad
  & \operatorname{ch}_{\omega_1} = & \ \frac{1}{2} \bigg[
    \frac{\eta(\tau)}{\vartheta_4(\mathfrak{b}_1)}
    - \frac{\eta(\tau)}{\vartheta_3(\mathfrak{b}_1)}
    \bigg] \ , \\
  \operatorname{ch}_{- \frac{1}{2}\omega_1} = & \ \frac{1}{2} \bigg[
    \frac{-i\eta(\tau)}{\vartheta_1(\mathfrak{b}_1)}
    + \frac{\eta(\tau)}{\vartheta_2(\mathfrak{b}_1)}
    \bigg], \qquad
  & \operatorname{ch}_{- \frac{3}{2}\omega_1} = & \ \frac{1}{2} \bigg[
    \frac{-i\eta(\tau)}{\vartheta_1(\mathfrak{b}_1)}
    - \frac{\eta(\tau)}{\vartheta_2(\mathfrak{b}_1)}
    \bigg] \ .
\end{align}
From the above expressions, it is obvious that under the spectral flow with $n_1 = 1,2,3$ (here the minus signs imply that the translation does not correspond to a proper map between states),
\begin{equation}
  \operatorname{ch}_0 \to \operatorname{ch}_{- \frac{1}{2}\omega_1}, \ - \operatorname{ch}_{\omega_1}, \ - \operatorname{ch}_{- \frac{3}{2}\omega_1} \ .
\end{equation}
Accordingly, under the flow, the $|\mathcal{N}_T\rangle$-equation (\ref{eq:NT-su2-1/2}) transforms into a weighted sum of equations (\ref{eq:NT-su2-1/2}), (\ref{eq:weight-5-su2-1/2}), (\ref{eq:weight-4-su2-1/2}), (\ref{eq:weight-3-su2-1/2}), (\ref{eq:weight-2-su2-1/2}),  corresponding to the null states $\frac{1}{\ell!}(h_1^1)^\ell|\mathcal{N}_T\rangle$, $\ell = 0, 1, 2, 3, 4$, respectively \footnote{For $\ell > 4$ the state $(h_1^1)^\ell|\mathcal{N}_T\rangle$ is identically zero, and does not correspond to a nontrivial equation.}.

In general, for all the Kac-Moody algebras considered previously, we can start with any null state $|\mathcal{N}\rangle$ that has a corresponding flavored MLDE. Then the translation $\sigma$
\begin{equation}
  \mathfrak{b}_i \to \mathfrak{b}_i + n_i \tau, \qquad
  \mathfrak{y} \to \mathfrak{y} + \sum_{i,j=1}^{r}K^{ij}n_j \mathfrak{b}_i
  + \sum_{i,j = 1}^{r} \frac{1}{2}K^{ij} n_i n_j \tau
\end{equation}
can be rewritten as an equivalent transformation at the level of null states,
\begin{align}
  |\mathcal{N}\rangle \xrightarrow{\sigma} \sum_{\ell \ge 0}\sum_{i_1\le ...\le i_\ell} \frac{(-1)^\ell}{\ell!} n_{i_1} ... n_{i_\ell} h^{i_1}_{1} \cdots h^{i_\ell}_{1} |\mathcal{N}\rangle \ .
\end{align}

\subsection{Twisted modules}

In the above, the translation was performed under the restriction that
\begin{equation}
  (\mathbf{n}^\vee, \alpha) \in \mathbb{Z}, \qquad \mathbf{n}^\vee \coloneqq \sum_{i = 1}^r n_i \alpha^\vee_i \ ,\qquad \forall \alpha \in \Delta \ .
\end{equation}
This condition ensures that all the twist parameters in the Eisenstein series remain unchanged after the translation,
\begin{equation}
  E_k \begin{bmatrix}
    +1 \\ \prod_{i = 1}^r b_i^{\lambda_i^\alpha}
  \end{bmatrix} \to \text{linear combination of } E_{k' \le k} \begin{bmatrix}
    + 1 \\ \prod_{i = 1}^r b_i^{\lambda_i^\alpha}
  \end{bmatrix} \ , \forall \alpha \in \Delta \ .
\end{equation}
This is because under the translation
\begin{equation}
  \prod_{i = 1}^r b_i^{\lambda_i^\alpha} \to 
  q^{\sum_{i = 1}^r n_i \lambda_i^\alpha} \prod_{i = 1}^r b_i^{\lambda_i^\alpha}, \qquad \text{where} \quad \sum_{i = 1}^r n_i \lambda_i^\alpha = (\mathbf{n}^\vee, \alpha) \ ,
\end{equation}
such that the power of $q$ is an integer.

We now show that, this restriction on the translation parameter $n$ can be modified to
\begin{equation}
  (\mathbf{n}^\vee, \alpha) \in \frac{1}{2}\mathbb{Z} \ .
\end{equation}
Such a translation leads to twisted modules and twisted flavored MLDEs of the Kac-Moody algebras.

Take $\widehat{\mathfrak{su}}(2)_k$ as examples. In \cite{2023arXiv230409681L}, twisted modules of admissible $\widehat{\mathfrak{su}}(2)_k$ algebras are discussed, whose characters can be obtained by performing an $\ell = -\frac{1}{2}$ translation/spectral-flow starting from untwisted admissible characters. As a trace over the module, the twisted character reads
\begin{equation}
  y^k b^{ - \frac{1}{2}k} q^{\frac{1}{16}k}\operatorname{tr} q^{L_0 - \frac{c}{24}} b_1^{h^1_0} q^{ - \frac{1}{4}h^1_0} \ .
\end{equation}
Compared with the untwisted character $y^k\operatorname{tr}q^{L_0 - c/24}b_1^{h^1_0}$, the operators $L_0, h_0^1$ are effectively translated to (here and below we will use the notation $\sigma^{-1/2}$ in \cite{2023arXiv230409681L} to denote the spectral flow by unit $-1/2$),
\begin{equation}
  \sigma^{- \frac{1}{2}}(L_0) = L_0 - \frac{1}{4}h^1_0 + \frac{1}{16}k, \quad
  \sigma^{-1/2}(h_0^1) = h_0^1 - \frac{1}{2}k \ .
\end{equation}
Measured by the new Hamiltonian $\sigma^{-1/2}(L_0)$ the conformal weight of $J^a_{-1}$ is tabulated in Table \ref{tab:spectral-flow-weights},
\begin{table}[h]
  \centering
  \begin{tabular}{c|ccc}
    & $h^1$ & $J^+$ & $J^-$ \\
    \hline
    $h$ & $1$ & $\frac{1}{2}$ & $\frac{3}{2}$
  \end{tabular}
  \caption{The weights of generators in $\mathfrak{su}(2)_k$ after $1/2$-unit spectral-flow. \label{tab:spectral-flow-weights}}
\end{table}
where we have omitted the universal translation due to the central element $k$.  Note that the difference in the conformal weights implies broken $SU(2)$ symmetry.

As pointed out in \cite{2023arXiv230409681L}, the unflavored twisted characters of an admissible algebra satisfy a corresponding unflavored twisted modular differential equation. Now we generalize the result to the flavored characters.

In the untwisted case, each flavored modular differential equation comes from a null state via Zhu's recursion, where the conformal weight $h[J^a] = 1$ of the generators $J^a_{-n}$ appearing in the state set the ``twist'' in all the Eisenstein series to $1$. Turning to the $(-\frac{1}{2})$-twisted modules, the change in the conformal weight precisely affects the twist in the Eisenstein seres in the flavored modular differential equations: starting with the equations for the non-twisted module, one simply replaces $E_k\big[\substack{1\\b_1^{\pm 2}}\big]$ by $E_k \big[\substack{-1 \\ b_1^{\pm 2}}\big]$, since such a factor must come from $J^\pm_k$ via Zhu's recursion, which now have half-integral conformal weights.

After the replacement, all the coefficients in the flavored MLDEs are analytic at $b_1 = 1$. This is because the singular coefficient must come from $E_1 \big[\substack{+1 \\ b_1^{\pm 2}}\big]$ in the untwisted equations, which is now replaced by $E_1 \big[\substack{-1 \\ b_1^2}\big]$, analytic at $b_1 = 1$. We conclude that the simultaneous solutions to the twisted flavored MLDEs, and hence all the twisted characters have finite $b_1 \to 1$ limit. In other words, the corresponding twisted modules are all ordinary. This conclusion is compatible with the theorem 5.2 in \cite{2023arXiv230409681L}.

\vspace{1em}
\textbf{Example: $\widehat{\mathfrak{su}}(2)_{-4/3}$}

Take $\widehat{\mathfrak{su}}(2)_{-4/3}$ as a simple example. This is a boundary admissible module with $k + 2 = \frac{2}{3}$, and also an entry in the Deligne-Cvitanovi\'{c} non-unitary series. The algebra coincides with the associated chiral algebra of the 4d $\mathcal{N} = 2$ Argyres-Douglas theory $(A_1, A_3)$ \cite{Cordova:2015nma,Buican:2015ina}. There are three admissible modules with the finite highest weight $0, - \frac{4}{3}\omega_1, - \frac{2}{3}\omega_1$, with the characters
\begin{align}
	\operatorname{ch}_0 = & \ y^{-4/3} \frac{\vartheta_1(2 \mathfrak{b}| 3\tau)}{\vartheta_1(2 \mathfrak{b} | \tau)} \ , \\
	\operatorname{ch}_1 = & \ y^{-4/3} b^{-2/3} q^{\frac{1}{6}} \frac{\vartheta_1(2 \mathfrak{b} - \tau| 3\tau)}{\vartheta_1(2 \mathfrak{b} | \tau)} \ , \\
	\operatorname{ch}_2 = & \ y^{-4/3} b^{- 4/3} q^{\frac{2}{3}} \frac{\vartheta_1(2 \mathfrak{b} - 2\tau| 3\tau)}{\vartheta_1(2 \mathfrak{b} | \tau)} \ .
\end{align}
All the admissible characters satisfy three flavored MLDEs,
\begin{equation}\label{su2Weight2}
	\left[
	D_q^{(1)} - \frac{3}{8} D_{b_1}^2
	- \frac{3}{2}E_1 \begin{bmatrix}
	  	+1 \\ b_1^2  
		\end{bmatrix}D_{b_1}
		+ 2E_2 \begin{bmatrix}
	  	+1 \\ b_1^2  
		\end{bmatrix}
		+ E_2 (\tau)
	\right] \operatorname{ch} = 0 \ ,
\end{equation}
\begin{equation}\label{su2Weight3}
	\left[D_{b_1}D_q^{(1)} - \left( 2 E_2 \begin{bmatrix}
		  	+1 \\ b_1^2  
			\end{bmatrix} - E_2(\tau) \right)D_{b_1}
			+ \frac{16}{3} E_3 \begin{bmatrix}
		  	+1 \\ b_1^2  
			\end{bmatrix}\right]\operatorname{ch} = 0 \ ,
\end{equation}
\begin{equation}\label{su2Weight4}
	\left[D_q^{(2)} + 2 E_3 \begin{bmatrix}
		  	+ 1 \\ b_1^2  
			\end{bmatrix} D_{b_1} - 8 E_4 \begin{bmatrix}
		  	+ 1 \\ b_1^2  
			\end{bmatrix} - 7 E_4(\tau)\right]\operatorname{ch} = 0 \ .
\end{equation}
As an example, we consider the character of the spectral-flowed vacuum module, whose character reads
\begin{align}
  \operatorname{ch}[\sigma^{-1/2}(M_0)] = - i q^{\frac{1}{24}} y_1^k b_1^{-1}b_1^{-\frac{1}{2}k} q^{\frac{1}{24}} \frac{\vartheta_1(2 \mathfrak{b}_1 - \frac{\tau}{2}| 3\tau)}{\vartheta_4(2 \mathfrak{b}_1|\tau)} \ .
\end{align}
Performing the replacement of the Eisenstein series $E_k\big[\substack{+1\\b_1^2}\big] \to E_k\big[\substack{-1\\b_1^2}\big]$, we arrive at the following twisted flavored modular differential equations,
\begin{equation}
	\left[
	D_q^{(1)} - \frac{3}{8} D_{b_1}^2
	- \frac{3}{2}E_1 \begin{bmatrix}
	  	-1 \\ b_1^2  
		\end{bmatrix}D_{b_1}
		+ 2E_2 \begin{bmatrix}
	  	-1 \\ b_1^2  
		\end{bmatrix}
		+ E_2 (\tau)
	\right] \operatorname{ch} = 0 \ ,
\end{equation}
\begin{equation}
	\left[D_{b_1}D_q^{(1)} - \left( 2 E_2 \begin{bmatrix}
		  	-1 \\ b_1^2  
			\end{bmatrix} - E_2(\tau) \right)D_{b_1}
			+ \frac{16}{3} E_3 \begin{bmatrix}
		  	-1 \\ b_1^2  
			\end{bmatrix}\right]\operatorname{ch} = 0 \ ,
\end{equation}
\begin{equation}\label{eq:NT-su2-4/3-twisted}
	\left[D_q^{(2)} + 2 E_3 \begin{bmatrix}
		  	-1 \\ b_1^2  
			\end{bmatrix} D_{b_1} - 8 E_4 \begin{bmatrix}
		  	-1 \\ b_1^2  
			\end{bmatrix} - 7 E_4(\tau)\right]\operatorname{ch} = 0 \ .
\end{equation}
It is easy to check that the spectral-flowed vacuum character $\operatorname{ch}[\sigma^{-1/2}M_0]$ indeed satisfies the above three twisted equations. In fact, all three $(-\frac{1}{2})$-unit spectral-flowed characters $\operatorname{ch}[\sigma^{- \frac{1}{2}}M_j]$, $j = 0, 1, 2$, satisfy these three equations. In particular, $\sigma^{-1/2}M_1$ is the module that corresponds to the canonical surface defect coupled to $(A_1, A_3)$ \cite{Cordova:2017mhb}. Note that when unflavored, $\sigma^{-\frac{1}{2}}M_0$ and $\sigma^{- \frac{1}{2}}M_2$ share the same unflavored characters, together with the unflavored character of $\sigma^{- \frac{1}{2}}M_1$ form the two unflavored solutions to the unflavored twisted equation \cite{2023arXiv230409681L}
\begin{equation}
  \biggl(D_q^{(2)} - 8 E_4 \begin{bmatrix}
    -1 \\ 1
  \end{bmatrix} - 7 E_4(\tau)\biggr) = 0 \ .
\end{equation}
This equation simply follows from equation (\ref{eq:NT-su2-4/3-twisted}) by sending $b_1 \to 1$, and recalling that $E_3 \big[\substack{-1 \\ 1}\big] = 0$.

Conversely, we can find all the simultaneous solutions to the above flavored MLDEs. Consider the ansatz in the twisted sector (the power $\sigma$ is not to be confused with the spectral flow symbol),
\begin{equation}
  \operatorname{ch} = q^{\sigma} \sum_{n = 0}^{+\infty} a_n(b_1) q^{n/2} \ ,
\end{equation}
To the leading order, there are three solutions
\begin{align}
  \sigma = & \ 0 , & a'_0 = & \ 0\\
  \sigma = & \ \frac{1}{6} , & a'_0 = & \ \frac{4}{9b_1}a_0 - b_1 a''_0 \ .
\end{align}
The first solution corresponds to the twisted module $\sigma^{-1/2}M_1$, while the second solution gives $a_0 = b_1^{+2/3}$ or $a_0 = b_1^{-2/3}$, corresponding to $\sigma^{-1/2}M_{0}, \sigma^{-1/2}M_2$, respectively. Thus we have recovered the full $\mathbb{Z}_2$-twisted spectrum of $\widehat{\mathfrak{su}}(2)_{-4/3}$ \cite{2023arXiv230409681L} by solving the flavored MLDEs.

\vspace{1em}
\textbf{Example: $\widehat{\mathfrak{su}}(2)_{-1/2}$}

Another simple example is $\widehat{\mathfrak{su}}(2)_{-1/2}$. Starting we (\ref{eq:NT-su2-1/2}), we replace the appropriate Eisenstein series to obtain the twisted flavored modular differential equations,
\begin{align}\label{eq:NT-su2-1/2-twisted}
  0 = & \ \Bigg[\frac{4}{5} \Bigg(10 E_2 \begin{bmatrix}
      -1\\b_1^2
  \end{bmatrix}^3+78 E_4 E_2 \begin{bmatrix}
      -1\\b_1^2
  \end{bmatrix}+40 E_1 \begin{bmatrix}
      -1\\b_1^2
  \end{bmatrix} E_3 \begin{bmatrix}
      -1\\b_1^2
  \end{bmatrix} E_2 \begin{bmatrix}
      -1\\b_1^2
  \end{bmatrix} \nonumber\\ 
  & \ +58 E_4 \begin{bmatrix}
      -1\\b_1^2
  \end{bmatrix} E_2 \begin{bmatrix}
      -1\\b_1^2
  \end{bmatrix}+216 \left(E_3 \begin{bmatrix}
      -1\\b_1^2
  \end{bmatrix}\right)^2 \nonumber\\
  & \ +E_2 \bigg(-10 E_2 \begin{bmatrix}
    -1\\b_1^2
  \end{bmatrix}^2
  -40 E_1 \begin{bmatrix}
      -1\\b_1^2
  \end{bmatrix} E_3 \begin{bmatrix}
      -1\\b_1^2
  \end{bmatrix}+68 E_4 \begin{bmatrix}
      -1\\b_1^2
  \end{bmatrix}+75 E_4\bigg) \nonumber\\
  \qquad& \ \qquad -126 E_6 \begin{bmatrix}
      -1\\b_1^2
  \end{bmatrix}
  -63 E_6\Bigg)\nonumber \\
  & \ +\Bigg(-20 E_3 \begin{bmatrix}
      -1\\b_1^2
  \end{bmatrix} E_1 \begin{bmatrix}
    -1\\b_1^2
  \end{bmatrix}^2-9 E_4 E_1 \begin{bmatrix}
      -1\\b_1^2
  \end{bmatrix}- \Big(5 E_2 \begin{bmatrix}
    -1\\b_1^2
  \end{bmatrix}^2+59 E_4 \begin{bmatrix}
      -1\\b_1^2
  \end{bmatrix} \Big) E_1 \begin{bmatrix}
      -1\\b_1^2
  \end{bmatrix} \nonumber \\
  & \ \qquad\quad -36 E_2 E_3 \begin{bmatrix}
      -1\\b_1^2
  \end{bmatrix}-83 E_2 \begin{bmatrix}
      -1\\b_1^2
  \end{bmatrix} E_3 \begin{bmatrix}
      -1\\b_1^2
  \end{bmatrix}-\frac{139}{5} \text{EE}(5) \begin{bmatrix}
      -1\\b_1^2
  \end{bmatrix}\Bigg) D_b \nonumber\\
  & \ +\left(\frac{5}{2} \left(E_2 \begin{bmatrix}
      -1\\b_1^2
  \end{bmatrix}\right)^2+10 E_1 \begin{bmatrix}
      -1\\b_1^2
  \end{bmatrix} E_3 \begin{bmatrix}
      -1\\b_1^2
  \end{bmatrix}-17 E_4 \begin{bmatrix}
      -1\\b_1^2
  \end{bmatrix}-\frac{75 E_4}{4}\right)D_b^2 \nonumber \\
  & \ +\frac{15}{4} E_3 \begin{bmatrix}
      -1\\b_1^2
  \end{bmatrix}D_b^3 -10 E_4 D_q^{(1)} + D_q^{(3)} \Bigg]\operatorname{ch} \ .
  \end{align}
  
With the same replacement of twist parameter, we obtain from (\ref{eq:weight-5-su2-1/2}), (\ref{eq:weight-4-su2-1/2}), (\ref{eq:weight-3-su2-1/2}), (\ref{eq:weight-2-su2-1/2}) the twisted flavored MLDEs of weight-5, 4, 3, 2, respectively. For later convenience, we write down the weight-four equation,
\begin{align}\label{eq:weight-4-su2-1/2-twisted}
  \Bigg[& - \frac{3}{2}D_q^{(2)} + 3 D_{b_1}^2 D_q^{(1)} - 3 E_2(\tau) D_q^{(1)}
  - \frac{3}{4} E_1 \begin{bmatrix}
    - 1 \\ b_1^2
  \end{bmatrix} D_{b_1}^3 \nonumber\\
  & \ + \bigg(6E_2(\tau) - 3 E_1 \begin{bmatrix}
    - 1 \\ b_1^2
  \end{bmatrix}^2
  - \frac{63}{8}E_2 \begin{bmatrix}
    - 1 \\ b_1^2
  \end{bmatrix} \bigg) D_{b_1}^2 \nonumber \\
  & \ + \bigg(\frac{9}{4}E_2(\tau) E_1 \begin{bmatrix}
    - 1 \\ b_1^2
  \end{bmatrix}
  + 6 E_1 \begin{bmatrix}
    - 1 \\ b_1^2
  \end{bmatrix}^3
  + 27 E_1 \begin{bmatrix}
    - 1 \\ b_1^2
  \end{bmatrix}E_2 \begin{bmatrix}
    - 1 \\ b_1^2
  \end{bmatrix}
  + \frac{387}{8}E_3 \begin{bmatrix}
    - 1 \\ b_1^2
  \end{bmatrix}
  \bigg) D_{b_1} \nonumber\\
  & \ - \frac{1}{48}\biggl(
    144E_2(\tau)^2
    - 633 E_4(\tau)
    + 144 E_1 \begin{bmatrix}
      - 1 \\ b_1^2
    \end{bmatrix}^2E_2 \begin{bmatrix}
      - 1 \\ b_1^2
    \end{bmatrix}
    - 80 E_2 \begin{bmatrix}
      - 1 \\ b_1^2
    \end{bmatrix}
  \nonumber\\
  & \ \qquad \qquad - 18 E_2 (\tau) \biggl(
    8 E_1 \begin{bmatrix}
      - 1 \\ b_1^2
    \end{bmatrix}^2
    + 21 E_2 \begin{bmatrix}
      - 1 \\ b_1^2
    \end{bmatrix}\biggr) \nonumber\\
    & \ \qquad \qquad + 1240 E_1 \begin{bmatrix}
      - 1 \\ b_1^2
    \end{bmatrix}E_3 \begin{bmatrix}
      - 1 \\ b_1^2
    \end{bmatrix}
    + 2833 E_4 \begin{bmatrix}
      - 1 \\ b_1^2
    \end{bmatrix}\biggr) 
  \Bigg]\operatorname{ch} \ .
\end{align}
It is straightforward to check that the following four twisted module characters, which is the full $\mathbb{Z}_2$-twisted spectrum, are indeed their solutions,
\begin{align}
  \operatorname{ch}\left[\sigma^{-\frac{1}{2}}M_0\right]=& \ \frac{ b_1^{\frac{1}{4}} q^{-\frac{1}{32}}}{2}\left[\frac{\eta(q \tau)}{\vartheta_4\left(\mathfrak{b}_1 - \frac{1}{4}\tau\right)}+\frac{\eta(\tau)}{\vartheta_3\left(\mathfrak{b}_1 - \frac{1}{4}\tau\right)}\right] \ , \\
\operatorname{ch}\left[\sigma^{-\frac{1}{2}}M_{\omega_1}\right]=& \ \frac{ b_1^{\frac{1}{4}} q^{-\frac{1}{32}}}{2}\left[\frac{\eta(\tau)}{\vartheta_4\left(\mathfrak{b}_1 - \frac{1}{4}\tau\right)}-\frac{\eta(\tau)}{\vartheta_3\left(\mathfrak{b}_1 - \frac{1}{4}\tau\right)}\right] \ , \\
\operatorname{ch}\left[\sigma^{-\frac{1}{2}}M_{- \frac{1}{2}\omega_1}\right]=& \ \frac{ b_1^{\frac{1}{4}} q^{-\frac{1}{32}}}{2}\left[\frac{-i \eta(\tau)}{\vartheta_1\left(\mathfrak{b}_1 - \frac{1}{4}\tau\right)}+\frac{\eta(\tau)}{\vartheta_2\left(\mathfrak{b}_1 - \frac{1}{4}\tau\right)}\right] \ ,\\
\operatorname{ch}\left[\sigma^{-\frac{1}{2}}M_{- \frac{3}{2} \omega_1}\right]=& \ \frac{ b_1^{\frac{1}{4}} q^{-\frac{1}{32}}}{2}\left[\frac{-i \eta(\tau)}{\vartheta_1\left(\mathfrak{b}_1 - \frac{1}{4}\tau\right)}-\frac{\eta(\tau)}{\vartheta_2\left(\mathfrak{b}_1 - \frac{1}{4}\tau\right)}\right] \ .
\end{align}
Conversely, solving the flavored MLDEs again recovers precisely the above four twisted characters. To leading order, with the ansatz $\operatorname{ch} = q^\sigma \sum_{n = 0}^{+\infty}c_n(b_1)q^{n/2}$, we find the constraints,
\begin{align}
  \sigma = & \ \frac{1}{96} , & c_0'' = & \ \frac{1}{16 b_1^2}c_0 - \frac{1}{b_1}c_0'\ , \\
  \text{or}, \qquad \sigma = & \ \frac{25}{96}, & c_0'' = & \ \frac{25}{16 b_1^2}c_0 - \frac{1}{b_1}c_0' \ .
\end{align}
Simple calculation reveals that the first line corresponds to $\sigma^{-1/2}M_0$ and $\sigma^{-1/2}M_{- \frac{1}{2}\omega_1}$, while the second line to $\sigma^{- 1/2}M_{\omega_1}$ and $\sigma^{- 1/2}M_{- \frac{3}{2}\omega_1}$. As a result, the twisted flavored MLDEs recovers the full $\mathbb{Z}_2$-twisted spectrum of $\widehat{\mathfrak{su}}(2)_{-1/2}$ \cite{2023arXiv230409681L}.

Now we consider the unflavoring limit. Before the $\mathbb{Z}_2$-twist, only the $|\mathcal{N}_T\rangle$-equation has a well-defined unflavoring limit as a unflavored MLDE, since all the other equations contain $D_{b_1}\operatorname{ch}$ or $D_{b_1}^3\operatorname{ch}$ terms with coefficients that are non-zero in the $b_1$ limit: $D_{b_1}\operatorname{ch}$ and $D_{b_1}^3 \operatorname{ch}$ themselves cannot be written as combinations of $D_q^{(n)}\operatorname{ch}$ and unflavored Eisenstein series. As a result, in the untwisted sector, the lowest order of unflavored MLDE is 3. However, the situation changes after the twist: the weight-four twisted flavored MLDE (\ref{eq:weight-4-su2-1/2-twisted}) actually has a non-trivial unflavoring limit as an unflavored MLDE. This can be seen by the following replacement in the $b_1 \to 1$ limit,
\begin{align}
  D_{b_1}^2 \operatorname{ch} \to & \ 6 D_q^{(1)} \operatorname{ch} + E_2(\tau)\operatorname{ch} + E_2 \begin{bmatrix}
    -1 \\ 1
  \end{bmatrix}\operatorname{ch} \ , \\
  D_q^{(1)}D_q^{(1)}\operatorname{ch} \to & \ D_q^{(2)}\operatorname{ch} - 2E_2(\tau)D_q^{(1)}\operatorname{ch} \ ,\\
  D_q^{(1)}E_2 \begin{bmatrix}
    -1 \\ 1
  \end{bmatrix} \to & \ 8 E_4 \begin{bmatrix}
    -1 \\ 1
  \end{bmatrix} + 2 E_4(\tau) - 2E_2(\tau) E_2 \begin{bmatrix}
    -1 \\ 1
  \end{bmatrix} \ ,\\
  D_q^{(1)}E_2(\tau) \to & \ 5 E_4(\tau) - E_2(\tau)^2 \ , \qquad
  E_\text{odd $k$}\begin{bmatrix}
    -1 \\ 1
  \end{bmatrix} \to 0 \ .
\end{align}
The first replacement comes from the Sugawara equation, and the second is simply the definition of $D_q^{(N)}$. All the $D_{b_1}^\text{odd}$ terms are multiplied with $E_\text{odd}\big[\substack{-1 \\ b_1^2}\big]$ which vanishes in the $b_1 \to 1$ limit. In the end, in the twisted sector, the lowest order of the unflavored MLDEs is 2 instead of 3, implying only two independent twisted ordinary characters. Explicitly, the second order equation reads \cite{2023arXiv230409681L}
\begin{equation}
 0 = \left({ D_q^{(2)} - \frac{5}{2} E_2 \begin{bmatrix}
    -1 \\ 1
  \end{bmatrix}D_q^{(1)}
  + \frac{643}{264} E_4(\tau) - \frac{169}{198}E_2 \begin{bmatrix}
    -1 \\ 1
  \end{bmatrix}^2 - \frac{529}{792}E_4 \begin{bmatrix}
    -1 \\ 1
  \end{bmatrix}}\right)\operatorname{ch} \ .
\end{equation}
In the unflavoring limit, $\operatorname{ch}[\sigma^{-1/2}M_0] = \operatorname{ch}[\sigma^{-1/2}M_{-\frac{1}{2}\omega_1}]$, $\operatorname{ch}[\sigma^{-1/2}M_{\frac{1}{2}\omega_1}] = \operatorname{ch}[\sigma^{-1/2}M_{- \frac{3}{2}\omega_1}]$.

\vspace{1em}
\textbf{Example: $\widehat{\mathfrak{su}}(3)_{-3/2}$}

As another example, we consider the algebra $\widehat{\mathfrak{su}}(3)_{-3/2}$. This is a boundary admissible Kac-Moody algebra in the Deligne-Cvitanovi\'c series. It is also the associated chiral algebra of the theory $(A_1, D_4)$, which has an $SU(3)$ flavor symmetry. The algebra has four admissible modules  with the affine highest weights \cite{Kac:1988qc,2016arXiv161207423K,2007arXiv0707.4129P,2012arXiv1207.4857A}
\begin{align}
	- \frac{3}{2} \widehat{\omega}_0, \qquad
	- \frac{3}{2} \widehat{\omega}_1, \qquad
	- \frac{3}{2} \widehat{\omega}_2, \qquad
	- \frac{1}{2} (\widehat{\omega}_0 + \widehat{\omega}_1 + \widehat{\omega}_2) \ .
\end{align}
The admissible characters can be written in closed form, given by respectively \cite{Kac:1988qc,2016arXiv161207423K}
\begin{align}
	\operatorname{ch}_{0} = & \ y^{-3/2} \frac{\eta(\tau)}{\eta(2 \tau)} \frac{
		\vartheta_1(\mathfrak{b}_1 - 2 \mathfrak{b}_2 |2 \tau)
		\vartheta_1( - \mathfrak{b}_1 - \mathfrak{b}_2 |2 \tau)
		\vartheta_1(-2 \mathfrak{b}_1 + \mathfrak{b}_2 |2 \tau)
	}{
		\vartheta_1(\mathfrak{b}_1 - 2 \mathfrak{b}_2 | \tau)
		\vartheta_1( - \mathfrak{b}_1 - \mathfrak{b}_2 | \tau)
		\vartheta_1(-2 \mathfrak{b}_1 + \mathfrak{b}_2 | \tau)
	} \ ,\\
	\operatorname{ch}_{1} = & \ - y^{-3/2} \frac{\eta(\tau)}{\eta(2 \tau)} \frac{
		\vartheta_4(\mathfrak{b}_1 - 2 \mathfrak{b}_2 |2 \tau)
		\vartheta_4( - \mathfrak{b}_1 - \mathfrak{b}_2 |2 \tau)
		\vartheta_1(-2 \mathfrak{b}_1 + \mathfrak{b}_2 |2 \tau)
	}{
		\vartheta_1(\mathfrak{b}_1 - 2 \mathfrak{b}_2 | \tau)
		\vartheta_1( - \mathfrak{b}_1 - \mathfrak{b}_2 | \tau)
		\vartheta_1(-2 \mathfrak{b}_1 + \mathfrak{b}_2 | \tau)
	} \ , \\
	\operatorname{ch}_{2} = & \ y^{-3/2} \frac{\eta(\tau)}{\eta(2 \tau)} \frac{
		\vartheta_1(\mathfrak{b}_1 - 2 \mathfrak{b}_2 |2 \tau)
		\vartheta_1( - \mathfrak{b}_1 - \mathfrak{b}_2 |2 \tau)
		\vartheta_1(-2 \mathfrak{b}_1 + \mathfrak{b}_2 |2 \tau)
	}{
		\vartheta_1(\mathfrak{b}_1 - 2 \mathfrak{b}_2 | \tau)
		\vartheta_1( - \mathfrak{b}_1 - \mathfrak{b}_2 | \tau)
		\vartheta_1(-2 \mathfrak{b}_1 + \mathfrak{b}_2 | \tau)
	} \ , \\
	\operatorname{ch}_3 = & \ y^{-3/2} \frac{\eta(\tau)}{\eta(2\tau)} \frac{
		\vartheta_4(\mathfrak{b}_1 - 2 \mathfrak{b}_2|2\tau)
	  \vartheta_4(-\mathfrak{b}_1 - \mathfrak{b}_2|2\tau)
	  \vartheta_4(-2\mathfrak{b}_1 + \mathfrak{b}_2|2\tau)
	}{
	  \vartheta_1(- \mathfrak{b}_1 - \mathfrak{b}_2|2\tau)
	  \vartheta_1(\mathfrak{b}_1 - 2 \mathfrak{b}_2|2\tau)
	  \vartheta_1(-2\mathfrak{b}_1 + \mathfrak{b}_2|2\tau)
	} \ .
\end{align}

Let us consider a particular translation acting on the vacuum module with 
\begin{equation}
  n_1 = - \frac{1}{6}, \qquad n_2 = - \frac{1}{3} \ .
\end{equation}
With this $n$, we have
\begin{center}
  \begin{tabular}{c|c|c}
    $\alpha$ & $\pm \alpha_1$ & $\pm \alpha_2, \pm(\alpha_1 + \alpha_2)$\\
    \hline
    $(\mathbf{n}^\vee, \alpha)$ & $0$ & $\pm \frac{1}{2}$
  \end{tabular}
\end{center}
The vacuum character is translated to
\begin{equation}
  \operatorname{ch}' = b_2^{\frac{3}{4}}q^{- \frac{1}{8}} \frac{\eta(\tau)}{\eta(2\tau)} \frac{
    \vartheta_1(\mathfrak{b}_1 - 2 \mathfrak{b}_2 + \frac{\tau}{2}|2\tau)
    \vartheta_1(2\mathfrak{b}_1 - \mathfrak{b}_2 |2\tau)
    \vartheta_1(\mathfrak{b}_1 + \mathfrak{b}_2 - \frac{\tau}{2}|2\tau)
  }{
    \vartheta_1(\mathfrak{b}_1 - 2 \mathfrak{b}_2 + \frac{\tau}{2}|\tau)
    \vartheta_1(2\mathfrak{b}_1 - \mathfrak{b}_2 |\tau)
    \vartheta_1(\mathfrak{b}_1 + \mathfrak{b}_2 - \frac{\tau}{2}|\tau)
  } \ .
\end{equation}
This is precisely one twisted character considered in \cite{2023arXiv230409681L}. Here, we can immediately write down several twisted flavored MLDEs that annihilate the above twisted character,
{\small\begin{align}
  0 = & \ D_q^{{(2)}} \operatorname{ch} + \Bigl(\frac{c}{2} + 3kr \Bigr)E_4(\tau)\operatorname{ch} +  \sum_{\alpha \in \Delta}\sum_{i = 1}^{r}K_{\alpha, -\alpha} f^{\alpha, -\alpha}{_i}E_3 \begin{bmatrix}
    \pm_\alpha \\ b^\alpha
  \end{bmatrix} D_{b_i}\operatorname{ch} 
  + 3k \sum_{\alpha \in \Delta} E_4 \begin{bmatrix}
    \pm_\alpha \\ b^\alpha
  \end{bmatrix} \operatorname{ch} \ , \nonumber
\end{align}}
{\small\begin{align}
  0 = & \ \Big(2\sum_i \mathfrak{b}_i D_{b_i}(D_q^{(1)} + E_2)
  -   \sum_{\alpha, i} \frac{|\alpha_i|^2}{2} m_i^\alpha  E_2 \begin{bmatrix}
    \pm_\alpha \\ b^\alpha  
  \end{bmatrix}(\alpha, \mathfrak{b}) D_{b_i}   - 2 k   \sum_{\alpha} (\alpha, \mathfrak{b}) E_3 \begin{bmatrix}
    \pm_\alpha \\ b^\alpha  
  \end{bmatrix} \Big) \operatorname{ch} \ ,
\end{align}}
{\small\begin{align}\label{eqSugawara}
  0 = \Bigg( 3D_q^{(1)}
  - K_{ij} D_{b_i}D_{b_j} 
  - \sum_{\alpha} K_{\alpha, - \alpha} & \ f^{\alpha, -\alpha}{_i} E_1 \begin{bmatrix}
      \pm_\alpha \\ b^\alpha
    \end{bmatrix}D_{b_i} - k r E_2(\tau)
  - k \sum_{\alpha} E_2 \begin{bmatrix}
      \pm_\alpha \\ b^\alpha
    \end{bmatrix} \Bigg) \operatorname{ch} \ , \nonumber
\end{align}}
where $r = 2$, $h^\vee = 3$, $c = -8$, and
\begin{equation}
  \pm_\alpha = e^{2\pi i (\mathbf{n}^\vee, \alpha)} = \left\{
    \begin{array}{cc}
      - 1 & \alpha \ne \pm \alpha_1\\
      + 1 & \alpha =  \pm \alpha_1
    \end{array}
  \right. \ .
\end{equation}
The sign change of the twist parameter can be read off by computing the scalar product $(\mathbf{n}^\vee, \alpha)$: the roots $\alpha$ such that the product is a half-integer should have the twist parameter change sign. The same spectral flow acting on the remaining untwisted admissible characters gives rise to further solutions to these flavored MLDEs. In the unflavoring limit, the twisted $|\mathcal{N}_T\rangle$-equation reduces to an unflavored MLDE \cite{2023arXiv230409681L},
\begin{equation}
  \biggl(D_q^{(2)} - 22 E_4(\tau) - 18 E_4 \begin{bmatrix}
    -1 \\ 1
  \end{bmatrix}\biggr)\operatorname{ch} = 0 \ .
\end{equation}
This suggests only two ordinary module characters, the flowed $\operatorname{ch}_0$ and $\operatorname{ch}_1$, in this particular twisted sector with the chosen $n$, while the flowed $\operatorname{ch}_2, \operatorname{ch}_3$ remain non-ordinary characters. Generalizing to translations with different $n$ is straightforward, where the corresponding twisted flavored MLDEs are those in the untwisted sector replacing all twist parameters by $e^{2\pi i (\mathbf{n}^\vee, \alpha)}$. 

\vspace{1em}
Finally, let us comment on the modular transformation and translation acting on the null states in the twisted sector. Just like in the untwisted case, the twisted flavored modular differential equations of admissible $\widehat{\mathfrak{su}}(2)_k$ Kac-Moody algebras enjoy quasi-modularity with respect to $\Gamma^0(2)$ (instead of $SL(2, \mathbb{Z})$). In other words, under the $STS$ transformation, an equation (corresponding to $|\mathcal{N}\rangle$) of higher weight transforms into a sum of equations of equal or lower weight, corresponding to null states $|\mathcal{N}\rangle, h^1_1 |\mathcal{N}\rangle, (h^1_1)^2 |\mathcal{N}\rangle$, etc. The precise form of the action is identical to the $STS$ on $|\mathcal{N}\rangle$ in the untwisted sector, whose explicit form will not be spelled out here.

Similarly, we summarize the translation property of the twisted flavored modular differential equations. For the $\widehat{\mathfrak{su}}(2)_k$ algebras, under the translation
\begin{equation}
  \mathfrak{b}_1 \to \mathfrak{b}_1 + n_1 \tau, \qquad \mathfrak{y} \to \mathfrak{y} + 2n_1 \mathfrak{b}_1 + n_1^2 \tau\ , \qquad n_1 \in \frac{1}{4}\mathbb{Z} \ ,
\end{equation}
the effective translation on the null states can be summarized into a universal simple formula
\begin{equation}
  |\mathcal{N}\rangle \to \sum_{\ell \ge 0} \frac{(-1)^\ell}{\ell!} n_1^\ell (h_1^1)^\ell|\mathcal{N}\rangle \ .
\end{equation}
For more general algebras, the translation can be effectively written in the universal formula, same for both untwisted and twisted sector,
\begin{align}
  |\mathcal{N}\rangle \xrightarrow{\sigma} \sum_{\ell \ge 0}\sum_{i_1\le ...\le i_\ell} \frac{(-1)^\ell}{\ell!} n_{i_1} ... n_{i_\ell} h^{i_1}_{1} \cdots h^{i_\ell}_{1} |\mathcal{N}\rangle \ .
\end{align}

\section*{Acknowledgments}
The authors would like to thank Tomoyuki Arakawa, Jin Chen, Bohan Li, Hao Li, Xin Wang, Yufan Wang, Wenbin Yan for useful discussions. Y.P. is supported by the National Natural Science Foundation of China (NSFC) under Grant No. 11905301. C. Z. is supported by Sun Yat-sen University Training Program of Research for Undergraduates, No. 20241764.


\appendix

\section{Special functions}

In this appendix we collect some useful properties of the special functions used in the main text.

The Jacobi theta functions are defined as infinite products using the $q$-Pochhammer symbol $(z;q) \coloneqq \prod_{k = 0}^{+\infty}(1 - zq^k)$,
\begin{align}
	\vartheta_1(\mathfrak{z}|\tau) \coloneqq & \ - i z^{\frac{1}{2}}q^{\frac{1}{8}}(q;q)(zq;q)(z^{-1};q) \ ,
	& \vartheta_2(\mathfrak{z}|\tau) \coloneqq & \ z^{1/2} q^{1/8}(q;q)(- zq;q)( - z^{-1};q) \ ,  \nonumber\\
	\vartheta_4(\mathfrak{z}|\tau) \coloneqq & \ (q;q)(zq^{1/2};q)(zq^{-1/2};q) \ ,
	& \vartheta_3(\mathfrak{z}|\tau) \coloneqq & \ (q;q)(-zq^{1/2};q)( - z^{-1}q^{1/2};q) \ . \nonumber
\end{align}
We often omit $|\tau$ from the notation. We also employ the variable naming convention using the fraktur font, e.g.,
\begin{equation}
	a = e^{2\pi i \mathfrak{a}}, \qquad b = e^{2\pi i \mathfrak{b}}, \qquad y = e^{2\pi i \mathfrak{y}} \ , \qquad z = e^{2\pi i \mathfrak{z}} \ ,
\end{equation}
except for the standard notation $q = e^{2\pi i \tau}$ with $\tau$ in the upper half complex plane. We have the standard relation between Jacobi theta function and the Dedekind $\eta$-function,
\begin{equation}
  \vartheta'(0) = 2\pi \eta(\tau)^3 \ .
\end{equation}

The $\vartheta$ functions enjoy simple shift property,
\begin{align}
  \vartheta_{1,2}(\mathfrak{z} + 1) = & \ - \vartheta_{1,2}(\mathfrak{z}),  & \vartheta_{3,4}(\mathfrak{z} + 1) = & \ + \vartheta_{3,4}(\mathfrak{z}) , \\
	\vartheta_{1,4}(\mathfrak{z} + \tau) = & \ - \lambda \vartheta_{1,4}(\mathfrak{z}),
	& \vartheta_{2,3}(\mathfrak{z} + \tau) = & \ + \lambda \vartheta_{2,3}(\mathfrak{z}) \ ,
\end{align}
where $\lambda \coloneqq e^{- 2\pi i \mathfrak{z}} e^{- \pi i \tau}$.

\vspace{1em}
The (twisted) Eisenstein series $E_k\big[\substack{\phi \\ \theta} \big]$ is defined as a $q$-series
\begin{align}
  E_{k \ge 1}\left[\begin{matrix}
    \phi \\ \theta
  \end{matrix}\right] \coloneqq & \ - \frac{B_k(\lambda)}{k!}  \\
  & \ + \frac{1}{(k-1)!}\sum_{r \ge 0}' \frac{(r + \lambda)^{k - 1}\theta^{-1} q^{r + \lambda}}{1 - \theta^{-1}q^{r + \lambda}}
  + \frac{(-1)^k}{(k-1)!}\sum_{r \ge 1} \frac{(r - \lambda)^{k - 1}\theta q^{r - \lambda}}{1 - \theta q^{r - \lambda}} \ , \nonumber
\end{align}
where $\phi = e^{2\pi i \lambda}$ and $\theta$ are often referred to as the characteristics. In this paper we will call $\phi$ the twist parameter as later discussions of twisted modules relates to this parameter. We will also call $k$ the (modular) weight of the Eisenstein series, as it is tied to the transformation property of $E_k$ under $SL(2, \mathbb{Z})$. $B_k(x)$ denotes the $k$-th Bernoulli polynomial, and the prime $^\prime$ means that the term with $r = 0$ should be omitted when $\phi = \theta = 1$. We also define $E_0\big[\substack{\phi\\\theta}\big] = -1$. In the limit $\phi, \theta \to 1$, we recover the standard Eisenstein series $E_k$,
\begin{equation}
	E_{2n}\begin{bmatrix}
		+1 \\ +1
	\end{bmatrix} = E_{2n}(\tau), \quad
	E_{2n + 1 \ge 3} \begin{bmatrix}
		+1 \\ +1
	\end{bmatrix} = E_{2n + 1}(\tau) = 0, \quad
	E_1 \begin{bmatrix}
		+1 \\ z
	\end{bmatrix}
	= \frac{1}{2\pi i } \frac{\vartheta_1'(\mathfrak{z})}{\vartheta_1(\mathfrak{z})} \ . 
\end{equation}
Note that $E_1 \big[\substack{1 \\ z}\big]$ has a simple pole at $z = 1$ (or, $\mathfrak{z} = 0$), since $\vartheta_1(\mathfrak{z}\to 0) \sim 2 \pi \mathfrak{z}q^{1/8}$, but $\vartheta'_1(0)\ne 0$.

There are several useful properties. First of all, the Eisenstein series enjoy the symmetry property
\begin{align}\label{Eisenstein-symmetry}
  E_k\left[\begin{matrix}
    \pm 1 \\ z^{-1}
  \end{matrix}\right] = (-1)^k E_k\left[\begin{matrix}
    \pm 1 \\ z
  \end{matrix}\right] \ .
\end{align}
The twisted Eisenstein series of neighboring weights are related by
\begin{align}\label{EisensteinDerivative}
  q \partial_q E_k\left[\begin{matrix}
    \phi \\ b
  \end{matrix}
  \right] = (- k) b \partial_b E_{k + 1}\left[\begin{matrix}
    \phi \\ b
  \end{matrix}
  \right]\ .
\end{align}
Here we see again that $q\partial_q = D_q^{(1)}$ and $b \partial_b$ raises the modular weight one two and one unit, respectively. More explicitly,
\begin{align}
  \text{odd} ~ n: \quad 0 = z \partial_z E_n \begin{bmatrix}
    1 \\ z
  \end{bmatrix} +  & \ (n + 1)E_{n + 1}\begin{bmatrix}
    1 \\ z
  \end{bmatrix} + E_{n + 1}(\tau) \nonumber\\
  & \  + E_1 \begin{bmatrix}
    1 \\ z
  \end{bmatrix}E_n \begin{bmatrix}
    1 \\ z
  \end{bmatrix}
  - \sum_{k = 2}^{n - 1}E_k(\tau)E_{n + 1 - k} \begin{bmatrix}
    1 \\ z
  \end{bmatrix} \ , \\
  \text{even} ~ n: \quad 0 = z \partial_z E_n \begin{bmatrix}
    1 \\ z
  \end{bmatrix} +  & \ (n + 1)E_{n + 1}\begin{bmatrix}
    1 \\ z
  \end{bmatrix} - E_n(\tau) E_1\begin{bmatrix}
    1 \\ z
  \end{bmatrix} \nonumber\\
  & \  + E_1 \begin{bmatrix}
    1 \\ z
  \end{bmatrix}E_n \begin{bmatrix}
    1 \\ z
  \end{bmatrix}
  - \sum_{k = 2}^{n - 1}E_k(\tau)E_{n + 1 - k} \begin{bmatrix}
    1 \\ z
  \end{bmatrix} \ .
\end{align}

When shifting the argument $\mathfrak{z}$ of the Eisenstein series by half-integral or integral units of $\tau$, or equivalently, shifting $z$ by $q^{\frac{n}{2}}$, one has
\begin{align}\label{Eisenstein-half-shift}
  E_k\left[\begin{matrix}
    \pm 1\\ z q^{\frac{n}{2}}
  \end{matrix}\right]
  =
  \sum_{\ell = 0}^{k} \left(\frac{n}{2}\right)^\ell \frac{1}{\ell !}
  E_{k - \ell}\left[\begin{matrix}
    (-1)^n(\pm 1) \\ z
  \end{matrix}\right] \ , \qquad n \in \mathbb{Z} \ .
\end{align}
Therefore, shifting the fugacity $z$ by a half-integer power of $q$ changes the twist parameter. A simple consequence is that\footnote{In fact, these equalities remain true even after replacing $1$ by $e^{2\pi i \lambda}$ and $- 1$ by $e^{2\pi i (\lambda + \frac{1}{2})}$.}
\begin{align}\label{Eisenstein-shift-1}
  E_k\left[\begin{matrix}
    \pm 1 \\ zq^{\frac{1}{2}}
  \end{matrix}\right]
  - E_k\left[\begin{matrix}
    \pm 1 \\ zq^{ - \frac{1}{2}}
  \end{matrix}\right]
  = & \ \sum_{m = 0}^{\floor{\frac{k - 1}{2}}} \frac{1}{2^{2m}(2m+1)!}E_{k - 1 - 2m}\left[\begin{matrix}
    \mp 1\\z
  \end{matrix}\right] \ ,
\end{align}
or more generally
\begin{align}
  E_k\left[\begin{matrix}
    \pm 1 \\ zq^{\frac{1}{2} + n}
  \end{matrix}\right]
  - E_k\left[\begin{matrix}
    \pm 1 \\ zq^{ - \frac{1}{2} - n}
  \end{matrix}\right]
  = & \ 2\sum_{m = 0}^{\floor{\frac{k - 1}{2}}} \left(\frac{2n+1}{2}\right)^{2m + 1}\frac{1}{(2m+1)!}E_{k - 1 - 2m}\left[\begin{matrix}
    \mp 1\\z
  \end{matrix}\right] \ . \nonumber
\end{align}

The Eisenstein series transforms non-trivially under $SL(2, \mathbb{Z})$ generated by
\begin{align}
  S: \tau \to - \frac{1}{\tau}, \ \mathfrak{z} \to \frac{\mathfrak{z}}{\tau}, \qquad
  \qquad
  T: \tau \to \tau + 1 , \ \mathfrak{z} \to \mathfrak{z} \ .
\end{align}
Concretely, $E_k \big[\substack{\pm 1 \\ \pm z}\big]$ transform under $S$,
\begin{align}
  E_n \begin{bmatrix}
    +1 \\ +z
  \end{bmatrix} \xrightarrow{S} &
  \left(\frac{1}{2\pi i}\right)^n\left[\bigg(\sum_{k \ge 0}\frac{1}{k!}(- \log z)^k y^k\bigg)
  \bigg(\sum_{\ell \ge 0}(\log q)^\ell y^\ell E_\ell \begin{bmatrix}
    + 1 \\ z
  \end{bmatrix}\bigg)\right]_n\ ,\\
  E_n \begin{bmatrix}
    -1 \\ +z
  \end{bmatrix} \xrightarrow{S} &
  \left(\frac{1}{2\pi i}\right)^n\left[\bigg(\sum_{k \ge 0}\frac{1}{k!}(- \log z)^k y^k\bigg)
  \bigg(\sum_{\ell \ge 0}(\log q)^\ell y^\ell E_\ell \begin{bmatrix}
    + 1 \\ -z
  \end{bmatrix}\bigg)\right]_n\ ,\\
  E_n \begin{bmatrix}
    1 \\ -z
  \end{bmatrix} \xrightarrow{S} &
  \left(\frac{1}{2\pi i}\right)^n\left[\bigg(\sum_{k \ge 0}\frac{1}{k!}(- \log z)^k y^k\bigg)
  \bigg(\sum_{\ell \ge 0}(\log q)^\ell y^\ell E_\ell \begin{bmatrix}
    -1 \\ +z
  \end{bmatrix}\bigg)\right]_n\ ,\\
  E_n \begin{bmatrix}
    -1 \\ -z
  \end{bmatrix} \xrightarrow{S} &
  \left(\frac{1}{2\pi i}\right)^n\left[\bigg(\sum_{k \ge 0}\frac{1}{k!}(- \log z)^k y^k\bigg)
  \bigg(\sum_{\ell \ge 0}(\log q)^\ell y^\ell E_\ell \begin{bmatrix}
    -1 \\ -z
  \end{bmatrix}\bigg)\right]_n\ ,
\end{align}
where $[ \ldots ]_n$ extracts the coefficient of $y^n$. Under the $T$-action,
\begin{align}
  E_n \begin{bmatrix}
    + 1 \\ + z
  \end{bmatrix} \xrightarrow{T}& \ E_n \begin{bmatrix}
    + 1 \\ + z
  \end{bmatrix}, & 
  E_n \begin{bmatrix}
    - 1 \\ + z
  \end{bmatrix} \xrightarrow{T}& \
  E_n \begin{bmatrix}
    - 1 \\ - z
  \end{bmatrix} \\
  E_n \begin{bmatrix}
    + 1 \\ - z
  \end{bmatrix} \xrightarrow{T}& \ E_n \begin{bmatrix}
    + 1 \\ - z
  \end{bmatrix}, & 
  E_n \begin{bmatrix}
    - 1 \\ - z
  \end{bmatrix} \xrightarrow{T}& \ 
  E_n \begin{bmatrix}
    - 1 \\ + z
  \end{bmatrix} \ .
\end{align}
We may also combine the two and obtain that under $STS$,
\begin{align}
  E_n \begin{bmatrix}
    -1 \\ z
  \end{bmatrix} \xrightarrow{STS}
  \left(\frac{1}{2\pi i}\right)^n\left[\bigg(\sum_{k \ge 0}\frac{1}{k!}(- \log z)^k y^k\bigg)
  \bigg(\sum_{\ell \ge 0}(\log q - 2\pi i)^\ell y^\ell E_\ell \begin{bmatrix}
    -1 \\ +z
  \end{bmatrix}\bigg)\right]_n\ . \nonumber
\end{align}

\bibliographystyle{utphys2}

\bibliography{ref}

\end{document}